\documentclass[11pt,twoside,a4paper,pdf]{article}  
\usepackage{graphicx,color}
\usepackage{times}
\usepackage{xspace}
\usepackage{booktabs}
\usepackage[english]{babel}
\usepackage{subfig}
\usepackage{amssymb}
\usepackage{cite}
\usepackage{xspace}
\usepackage{a4wide}
\usepackage{authblk,setspace,caption3,babel}
\usepackage{amsmath,multicol}

\newcommand{\mrm}[1]{\ensuremath{\mathrm{#1}}\xspace}

\newcommand{\tsc}[1]{\textsc{#1}}

\newcommand{\EqRef}[1]{Eq.~(\ref{#1})\xspace}

\newcommand{\secRef}[1]{section~\ref{#1}\xspace}

\newcommand{\TabRef}[1]{Tab.~\ref{#1}\xspace}

\newcommand{\FigRef}[1]{Fig.~\ref{#1}\xspace}

\newcommand{\TabsRef}[1]{Tabs.~\ref{#1}\xspace}

\newcommand{\FigsRef}[1]{Figs.~\ref{#1}\xspace}
\newcommand{\inst}[1]{$^{#1}$}
\renewcommand{\and}{, }

\graphicspath{{images/Tuning/},{images/Results/}}

\begin{document}

\vspace*{-1.8cm}\begin{minipage}{\textwidth}
\flushright
CERN-PH-TH/2014-018\\
DESY-14-014\\
KA-TP-05-2014\\
MCNET-14-04\\
\end{minipage}
\vskip1.25cm
{\Large\bf
\begin{center}
Revisiting Radiation Patterns in $e^+e^-$ Collisions
\end{center}}
\vskip5mm
{\begin{center}
{\large 
N.~Fischer\inst{1}\and 
S.~Gieseke\inst{1}\and
S.~Pl\"atzer\inst{2}\and
P.~Skands\inst{3}
}\end{center}\vskip3mm
\begin{center}
\parbox{0.9\textwidth}{
\inst{1}: KIT, Institute for Theoretical Physics, D-76128 Karlsruhe, Germany\\
\inst{2}: DESY Theory Group, Notkestrasse 85, D-22607 Hamburg, Germany\\
\inst{3}: Theoretical Physics, CERN, CH-1211,
Geneva 23, Switzerland
}
\end{center}
\vskip5mm
\begin{center}
\parbox{0.85\textwidth}{
\begin{center}
\textbf{Abstract}
\end{center}\small
We propose four simple event-shape variables for semi-inclusive $e^+e^- \to
4$-jet events. The observables and cuts are designed to be especially
sensitive to subleading aspects of the event structure, and allow to
test the reliability of phenomenological QCD models in greater detail. Three of 
them, $\theta_{14}$, $\theta^*$, and $C_2^{(1/5)}$, focus on
soft emissions off three-jet topologies with a small opening angle,
for which coherence effects beyond the leading QCD dipole pattern are
expected to be enhanced. A complementary variable, $M_L^2/M_H^2$,
measures the ratio of the hemisphere masses in 4-jet events with a
compressed scale hierarchy (Durham $y_{23}\sim y_{34}$), for which
subleading $1\to 3$ splitting effects are expected to be enhanced. 
We consider several different
parton-shower models, spanning both conventional 
and dipole/antenna ones, all tuned to the same
$e^+e^-$ reference data, and show that a measurement of 
the proposed observables would allow
for additional significant discriminating power between the models. 
}
\end{center}
\vspace*{1cm}
\section{Introduction \label{sec:intro}}
General-purpose event generators
(see~\cite{Buckley:2011ms,Beringer:1900zz,Seymour:2013ega,Gieseke:2013eva} for recent reviews)  
aim to give a complete description
of high-energy interactions, down to the level of individual
particles. 
They are extensively used as research vessels for exploring new
approaches to phenomenological questions within and beyond the
Standard Model, and they are relied upon to provide explicit
simulations of high-energy reactions in a broad variety of contexts. 
The achievable accuracy depends both on the
inclusiveness of the chosen observable and on the  
sophistication of the calculation itself. An important driver for the
latter is obviously the development of improved theoretical models; but it  
also depends crucially on the available constraints on the remaining
free parameters. Using existing data to constrain these is referred to
as generator tuning. 

The main experimental reference for final-state radiation and
fragmentation studies is the process $e^+e^- \to Z/\gamma^* \to
\mrm{hadrons}$.  Prior to and during the LEP era, a large set of event
measurements were performed (see, e.g.,
\cite{Barate:1996fi,Abreu:1996na,Achard:2004sv,Akrawy:1990yx}) and
used to constrain the shower and hadronization models of the day, such
as \tsc{Herwig}~\cite{Corcella:2000bw},
\tsc{Jetset}/\tsc{Pythia}~\cite{Sjostrand:2006za}, and
\tsc{Ariadne}~\cite{Lonnblad:1992tz}.  Most of the relevant analyses
were corrected to the particle level and have subsequently been
encoded in \tsc{Rivet}~\cite{Buckley:2010ar}.  This makes it
straightforward to apply \emph{almost} the same comprehensive battery
of tests to any model today\footnote{With a few notable exceptions
  not available for direct
  hadron-level MC comparisons,
  like the four-jet
  angles~\cite{Abreu:1990ce,Decamp:1992ip,Abbiendi:2001qn},
  observables directly 
  sensitive to coherence~\cite{Abdallah:2004uu}, and 
  colour-reconnection
  constraints~\cite{Abbiendi:2003ri,Achard:2003pe,Achard:2003ik,Siebel:2005uw,Abbiendi:2005es,Schael:2006ns,Abdallah:2006uq}.}. 
 A main question we wish to
examine in this study is whether the existing
constraints are sufficient in the context of present-day models. The
reasons to ask this question are threefold.

Firstly, current parton-shower models are, in fact, quite
sophisticated, at least as far as pure final-state radiation effects
are concerned. For instance, they all include colour coherence (though
the way this is achieved differs from model to model), the inclusion
of dominant contributions of two-loop splitting kernels by suitable
renormalization-scale choices (e.g., $\mu_R\propto p_\perp$), and
effects of momentum conservation (again with individual models
employing different ``recoil'' strategies), and several even
incorporate further subleading aspects such as gluon-polarization or
helicity-conservation effects.  Their precision is therefore typically
much better than their nominal ``leading-logarithmic'' (LL) labels
indicate; in comparison with the experimental uncertainties at LEP,
differences on observables dominated by LL effects are typically too
small to show up clearly (cf., e.g.,~\cite{Karneyeu:2013aha}).  It is
therefore interesting to study whether more information can be
extracted from variables designed to remove LL contributions and
isolate specific subleading aspects.

Secondly, over the last decade, several completely new parton-shower models have
been 
formulated~\cite{Gieseke:2003rz,Sjostrand:2004ef,Nagy:2005aa,Krauss:2005re,Giele:2007di,Dinsdale:2007mf,Platzer:2009jq,Nagy:2012bt,Schumann:2007mg,Winter:2007ye},
in the context of a new generation of MC generators such as
\tsc{Herwig++}~\cite{Bahr:2008pv,Platzer:2009jq}, \tsc{Pythia~8}~\cite{Sjostrand:2007gs}, \tsc{Sherpa}~\cite{Gleisberg:2008ta},
and \tsc{Vincia}~\cite{Giele:2007di}.
Many of the new shower models build on the coherent QCD dipole-antenna
 formalism~\cite{Gustafson:1987rq,Catani:1996vz,Kosower:2003bh,GehrmannDeRidder:2005cm}
 and aim explicitly at facilitating combinations with higher-order matrix
elements~\cite{Nagy:2012bt,Platzer:2011bc,Giele:2011cb,Hartgring:2013jma}
 (so-called ``matching''). 
These models were not present 
during the main era of $ee$ measurements, and hence could not directly 
inform the selection of observables. Thus, it is natural at this point  
to reconsider whether there are additional interesting observables,
which could provide further non-trivial constraints on modern
generators. 

Thirdly, the desire for reliable descriptions of jet production and jet
substructure for signal and background estimates at the LHC is causing
the subleading aspects of shower models and matrix-element matching
strategies to come under increasing scrutiny, in particular in the
context of the interplay between matching and tuning. While all shower
and matching strategies are designed to have the same leading
behaviours, they do exhibit differences at subleading levels, making
subleading-sensitive observables especially interesting for cross
checks.

In this paper, we are interested mainly in inclusive four-jet
observables sensitive to coherence properties and to effective $1\to
3$ splittings. The starting points are the $\theta^*$ variable
proposed in~\cite{Platzer:2009jq}, $\theta_{14}$ and $M_L^2/M_H^2$
proposed in~\cite{AlcarazMaestre:2012vp}, and the energy correlation
functions proposed in~\cite{Larkoski:2013eya}. The former two,
$\theta^*$ and $\theta_{14}$, are designed to be sensitive to the
coherent emission of a soft fourth jet from a three-parton state (with
cuts restricting the opening angles of the jets, as will be described
below), with a radiation pattern dictated by colour coherence. In
particular, they can be used to test whether the angular distribution
of the fourth jet is well described by a three-parton system
represented by partons / dipoles / antennae, and how this description
depends upon the choice of shower ordering variable.  The latter two
variables, $M_L^2/M_H^2$ (the ratio of hemisphere masses) and the
energy correlation functions, have sensitivity to the effective
description of $1\to 3$ splittings and the energy spectrum of the
fourth jet, respectively, as will be discussed below. For
all observables, we impose an explicit cut on the Durham $k_T$
resolution scale of the fourth jet, $y_{34}>0.0045$ (corresponding to
$\ln(y_{34})~>~-5.4$), thus restricting it to be in the perturbative
domain and avoiding possible contamination from $B$ decays.

We examine six different parton-shower models: the default angular-ordered
parton shower of \tsc{Herwig++}~\cite{Gieseke:2003rz}, 
the $p_\perp$- and virtuality-ordered dipole showers of
\tsc{Herwig++}~\cite{Platzer:2009jq}, 
the default $p_\perp$-ordered shower of
\tsc{Pythia~8}~\cite{Sjostrand:2004ef}, and the $p_\perp$- and
$m_{\mrm{ant}}^2$-ordered antenna showers of
\tsc{Vincia}~\cite{Giele:2007di}. 

The salient properties of each shower model will be summarized briefly in
\secRef{sec:models}. As a cross check, and 
to ensure a fair comparison between the models, we tune all of them to
the same reference data in \secRef{sec:tuning}. The main study of 
soft-jet and event-shape variables is presented in
\secRef{sec:results}. Finally, we round off with conclusions in
\secRef{sec:conclusions}. 

\section{Theory Models \label{sec:models}}

Parton showers are not guaranteed to respect coherence. For example,
in a traditional shower based on the collinear DGLAP
formalism~\cite{Gribov:1972ri,Altarelli:1977zs,Dokshitzer:1977sg}, the
linear sum of $n$ DGLAP splitting kernels (one for each parton in an
$n$-parton state) can substantially overcount the amount of wide-angle
soft radiation in comparison, e.g.~\cite{Bengtsson:1986et}, with
$(n+1)$-parton matrix elements. 
Physically speaking, if we approximate the radiation from an
$n$-parton (``colour-multipole'') state by the incoherent sum of $n$
monopole terms, there is a substantial risk that highly important
destructive-interference effects will be neglected, leading to double
counting of soft gluon emission.

It was found in the early eighties~\cite{Marchesini:1983bm}, that
DGLAP-based parton showers can nonetheless be brought to agree with the
correct soft limits of QCD (up to azimuthal averaging effects), by choosing 
the shower ordering variable to be proportional to energy times
angle. This is the basis of the \emph{angular-ordered}
showers~\cite{Gieseke:2003rz} in
\tsc{Herwig++}, which is the first shower model we include 
in our study.

An alternative DGLAP-based shower model is that of
\tsc{Pythia~8}, the second model included in our study. 
In this framework~\cite{Sjostrand:2004ef}, small
opening angles are reinterpreted as corresponding to highly boosted
colour dipoles. The resulting Lorentz-boosted DGLAP radiation patterns
combined with an ordering in  transverse momentum of the dipoles are used to
obtain approximately coherent results. 

A more formal definition of showers based on colour dipoles can be
obtained by replacing the DGLAP splitting kernels by intrinsically
coherent radiation functions such as Catani-Seymour (CS) dipole
functions~\cite{Catani:1996vz} or QCD antenna functions (also called
Lund dipoles)~\cite{Gustafson:1987rq,Kosower:2003bh}. 
These reproduce the leading collinear \emph{and} soft singularities of
QCD amplitudes for each {\it single} emission without the need of a
particular phase-space restriction as present in angular-ordered
showers. They can, however, differ in the ordering variable, affecting
{\it multiple} emissions and hence potentially higher-order coherence
properties. Another difference is the recoil strategy taken, which can
lead to differences at the level of next-to-leading logarithms or
beyond. In order to explore these ambiguities more fully, we include four
different variants of dipole-antenna shower models in our study, two
based on a dipole formalism and two based on antennae, with
differences as follows.

For each radiation term, the dipole formalism identifies a single
parton as the \emph{emitter}, with a colour partner assigned to be the
\emph{spectator}. The recoil is constrained to be purely longitudinal,
in the rest frame of the dipole pair.  By itself, the \emph{dipole
  radiation function} only accounts for half of the soft singularity
of the dipole pair, and there is no collinear singularity associated
with the spectator.  There is a separate radiation term in which the
roles of the two are reversed, such that the sum is correct in all the
infrared limits.  The preferred choice of ordering variable is transverse
momentum, $p_{\perp\mrm{dip}}$, the relative transverse momentum of
the splitting products with collinear direction defined by the
spectator. This defines the third model included in this study. 
As a fourth option,
we consider ordering in the virtuality of the splitting products,
$q_\mrm{dip}$ (see table~\ref{tab:models} for precise definitions).

In the antenna formalism, there is no unique distinction between
emitters and spectators. Instead, a single \emph{antenna radiation
  function} captures the
collinear limits of both of the colour partners together with their
full soft singularity, and a $2\to 3$ kinematics map is used, which 
smoothly interpolates between the two collinear limits (both
parents generally acquire some recoil). In this context, 
it has been shown explicitly~\cite{Hartgring:2013jma} that the choice
of $p_\perp$ as evolution 
variable absorbs all logarithms through second order in $\alpha_s$
(i.e., up to and including $\alpha_s^2 \ln Q^2$ corrections), hence this
is the preferred choice, defining the fifth shower model
in our study. As an alternative, we also consider ordering in
antenna mass, which is known to exhibit an $\alpha_s^2 \ln Q^2$ discrepancy with
respect to second-order QCD~\cite{Hartgring:2013jma}. 

\begin{table}[tp]
\centering
\begin{tabular}{rp{2.5cm}llp{4.9cm}l}\toprule
 & & Radiation  & Kinematics         & Ordering variable\\
 & & functions  & (a.k.a.~recoils)   & for gluon emissions  \\
\midrule
\bf 1 & \tsc{Herwig++}
  & \noindent\centering DGLAP & Global & $\tilde q^2=\frac{Q_I^2M_{IK}^4}{Q_K^2(M_{IK}^2-Q_I^2-Q_K^2)}$
\\    & (default) & 
\\[1ex]
\bf 2 & \tsc{Pythia~8}
  & DGLAP  & Dipole & $p^2_{\perp\mrm{evol}} = \frac{Q_I^2(M^2_{IK}-Q^2_K)(Q^2_I+Q^2_K)}{(M^2_{IK}+Q^2_I)^2}$
\\    & (default) & & &
\\[1ex]
\bf 3 & \tsc{Herwig++}
  & Dipole  & Dipole & $p^2_{\perp\mrm{dip}} =  \frac{Q_I^2Q_K^2(M_{IK}^2-Q_I^2-Q_K^2)}{(M_{IK}^2-Q_I^2)^2}$
\\    & dipole $p_\perp^2$
  &    &        & 
\\[1ex]
\bf 4 & \tsc{Herwig++}
  &  Dipole & Dipole & $q_\mrm{dip}^2 = Q_I^2$ 
\\  & dipole virtuality & 
\\[1ex]
\bf 5 & \tsc{Vincia} \newline antenna $p_\perp^2$
  & Antenna & Antenna & $p_{\perp\mrm{ant}}^2 = \frac{Q^2_I Q^2_K}{M_{IK}^2}$ 
\\[4ex]
\bf 6 & \tsc{Vincia}
  & Antenna & Antenna & $m_{\mrm{ant}}^2 = \min(Q^2_I,Q^2_K)$
\\ &  antenna mass & & & 
\\[1ex]\bottomrule
\end{tabular}
\caption{The six shower models considered in this paper. The ordering
  variables shown correspond to $I\to ij$ for the DGLAP models, the
  same with $K$ as the spectator for the CS dipole models, and to $IK
  \to ijk$ for the antenna ones. We use the notation $Q_I^2 = (p_i +
  p_j)^2$, $Q_K^2 = (p_j+p_k)^2$, and $M_{IK}^2 = (p_I + p_K)^2 = (p_i
  + p_j + p_k)^2$.  The \tsc{Pythia~8} evolution
  variable is defined as $p_{\perp\mrm{evol}}^2 = z(1-z)Q_I^2$ with
  $z=(M^2_{IK}-Q^2_K)/(M^2_{IK}+Q^2_I)$ 
the fraction of the light-cone momentum of parton
  $I$ carried by parton $i$, in the DGLAP functions,
  $P(z)$. \label{tab:models}
}
\end{table}
A systematic comparison of the salient differences between these six different
shower models is given in \TabRef{tab:models}. Contours of constant
value of each of the corresponding evolution variables are shown in
\FigRef{fig:evolution}, over the triangular dipole branching phase
space. Labelling the pre- and post-branching partons by $IK \to ijk$,
the axes of the plots are defined by the dimensionless branching
invariants $Q_I^2/M_{IK}^2$ and
$Q_K^2/M_{IK}^2$, so that the collinear singularities 
lie along the axes and the soft singularity lies at (0,0). Note
that the DGLAP- and dipole-based 
evolution variables, 2--4, correspond to the evolution of a single parton,
$I$, hence the corresponding radiation functions only have collinear
singularities along the $y$ axis; the antenna evolution variables, 5--6, 
correspond to the evolution of the $IK$ antenna, 
with collinear singularities along both axes. 
\begin{figure}[tp]
\centering\begin{tabular}{ccc}
\includegraphics*[scale=0.42]{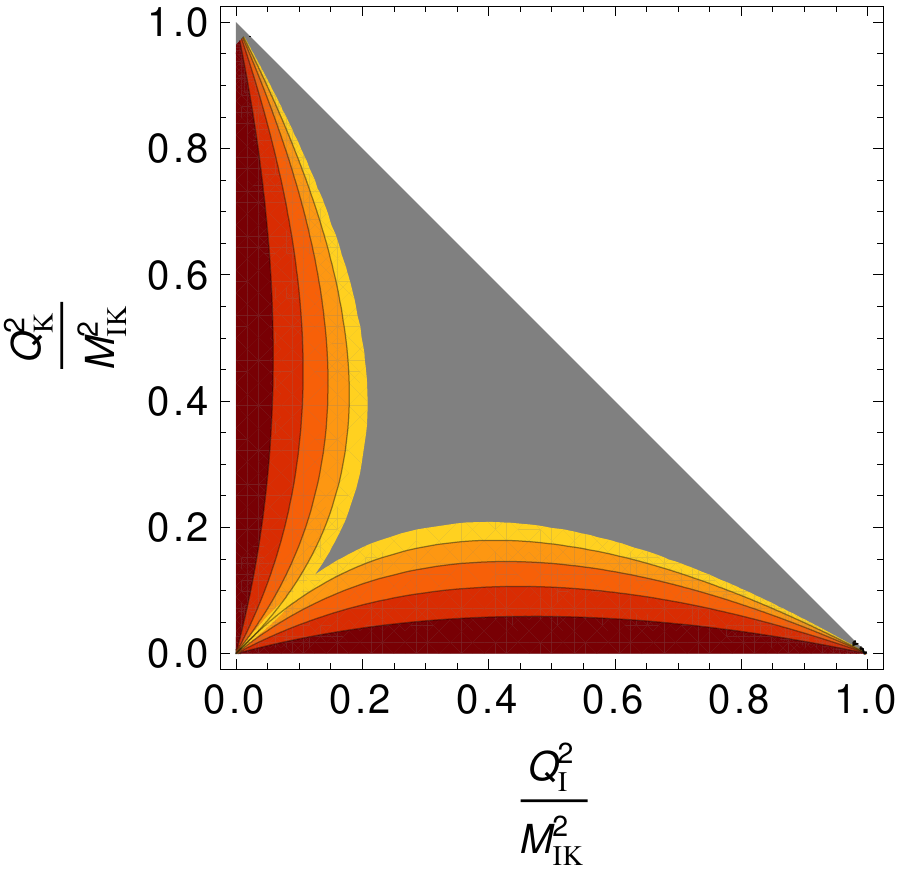} & 
\includegraphics*[scale=0.42]{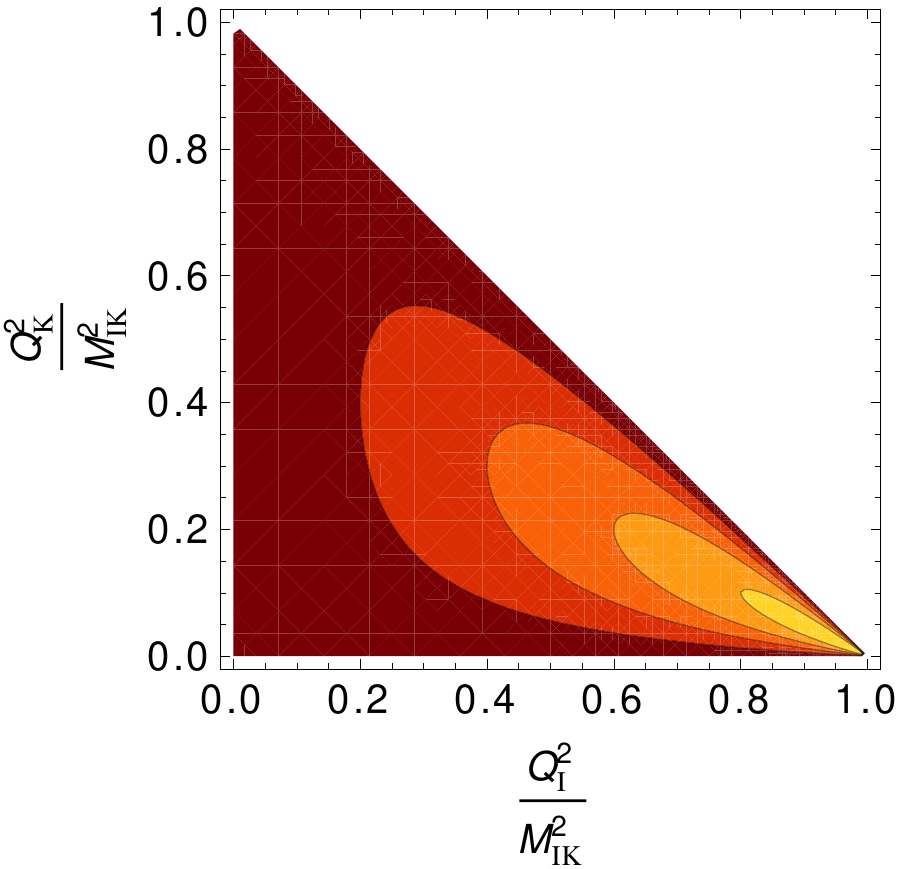} & 
\includegraphics*[scale=0.42]{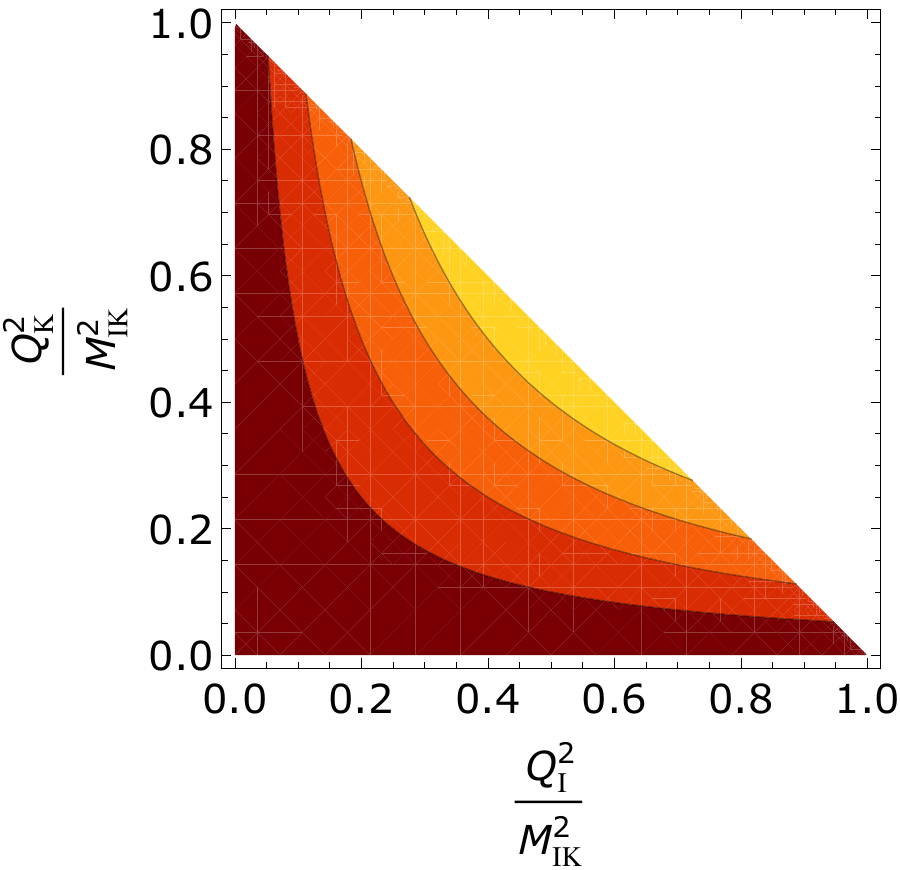} \\
1) \tsc{Herwig++} angle & 
3) \tsc{Herwig++} $p_{\perp \mrm{dip}}^2$ & 
5) \tsc{Vincia} $p_{\perp\mrm{ant}}^2$ \\[2ex]
\includegraphics*[scale=0.42]{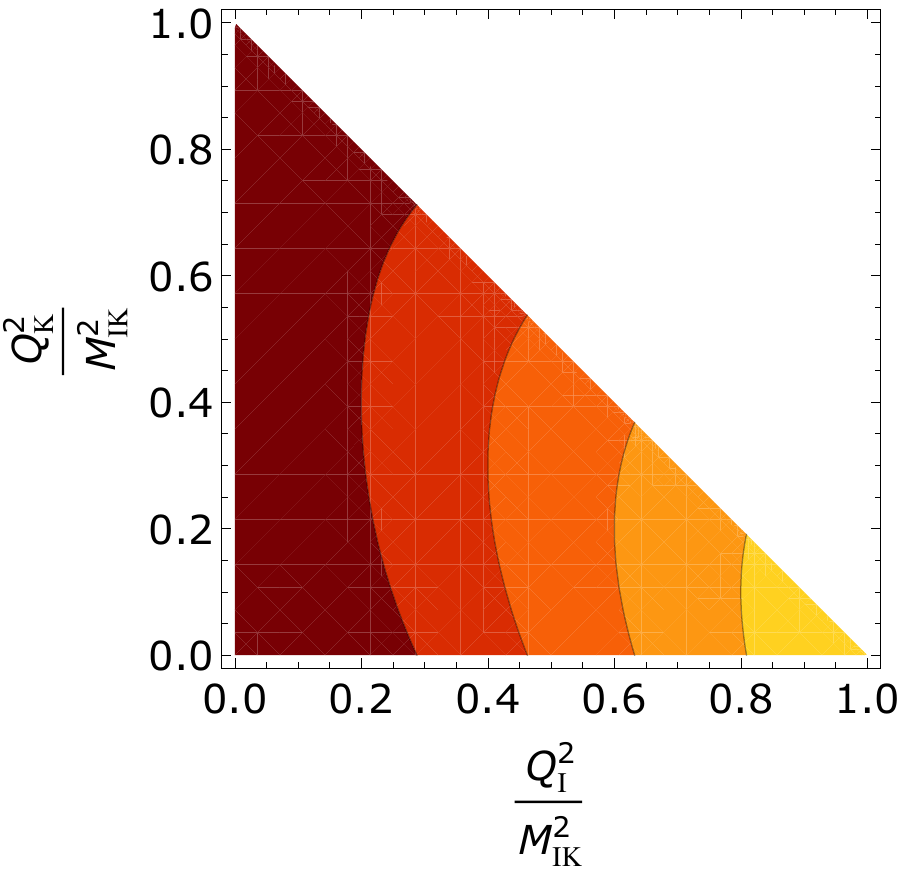} &
\includegraphics*[scale=0.42]{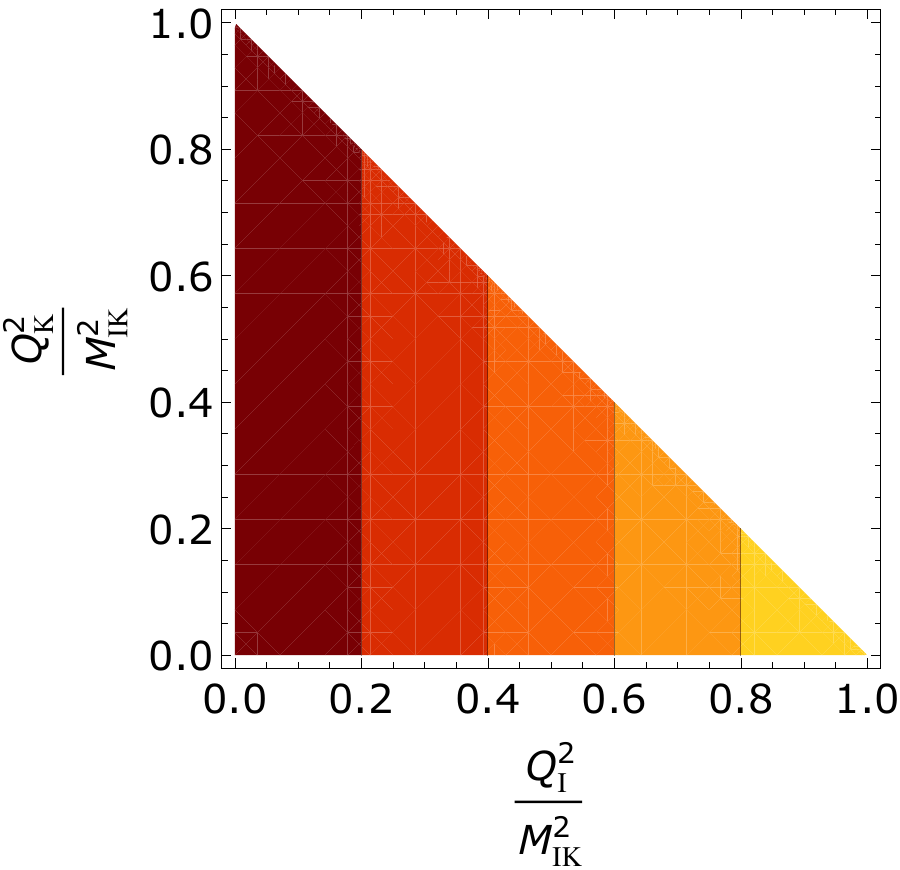} & 
\includegraphics*[scale=0.42]{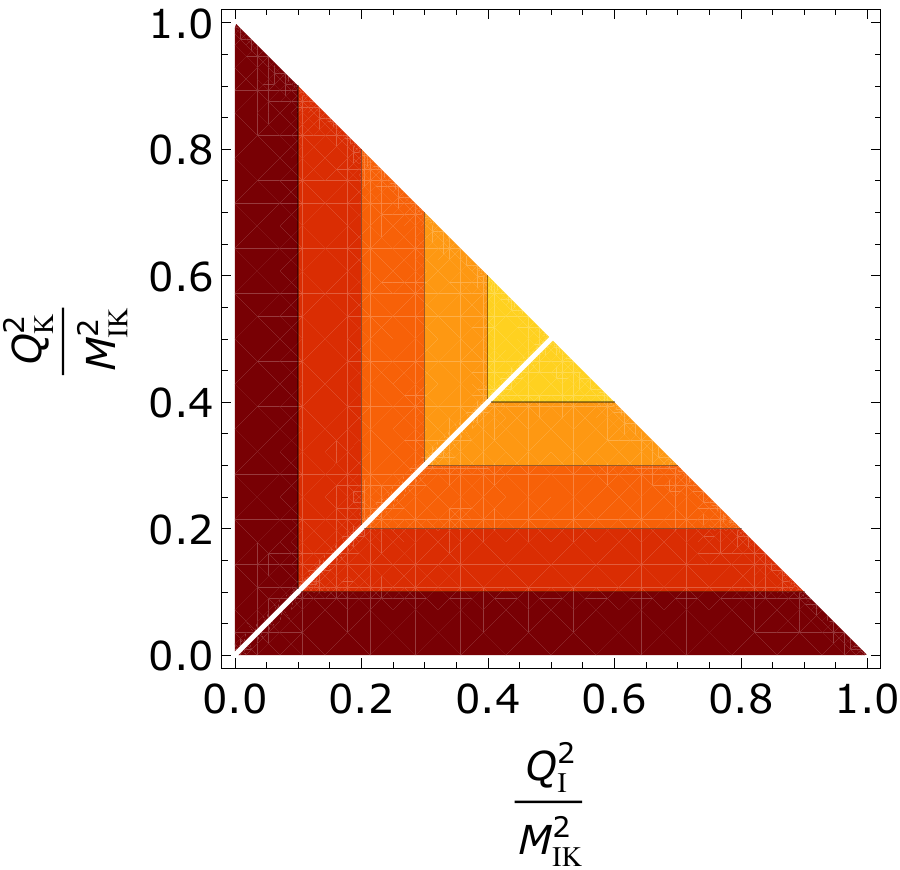} \\
2) \tsc{Pythia} $p_{\perp\mrm{evol}}^2$ & 
4) \tsc{Herwig++} $q_\mrm{dip}^2$ & 
6) \tsc{Vincia} $m_\mrm{ant}^2$ 
\end{tabular}
\caption{Illustration of the progression of the shower evolution
  variables over the dipole phase space, for each of the 
  models listed in \TabRef{tab:models}. Note that 2--4 correspond to
  radiation functions whose only singularities lie along the $y$
  axis, while 1, 5, and 6 have singularities along both the $x$ and
  $y$ axes. 
\label{fig:evolution}}
\end{figure}

In order to focus on the pure shower aspects and make the models more
directly comparable, a few non-default choices have been made in the
context of our study. In particular for \tsc{Vincia}, ME corrections
at both LO~\cite{Giele:2011cb} and NLO~\cite{Hartgring:2013jma} were
switched off, and we use the smoothly-ordered 
showers~\cite{Giele:2011cb} with a one-loop running of $\alpha_s$. 
The \tsc{Herwig++} dipole-shower simulations likewise used a one-loop running 
and no matrix-element corrections nor NLO matching has been applied. 
For the default shower models (angular-ordered in \tsc{Herwig++} and 
$p_{\perp\mrm{evol}}$-ordered in \tsc{Pythia}), we use the respective
default settings, which includes matrix-element corrections for the
first emission, for both codes, and one-loop (two-loop) running for
\tsc{Pythia} (\tsc{Herwig++}), respectively.
As a cross-check, we investigated the effect of including NLO matching for the 
$p_{\perp \mrm{dip}}$-ordered dipole shower of \tsc{Herwig++}
and found that the four-jet observables, which we study here,
are not sensitive to these corrections.
An enlarged set of results, including plots of the 
last-mentioned study and strong vs.\ smooth ordering,
will be included in~\cite{nadine}.

\section{Tuning \label{sec:tuning}} 

In order to compare the models on as equal a footing as possible, we
first adjust (``tune'') the shower and hadronization parameters of
each model to the same set of existing LEP measurements. We perform this 
tuning with the \tsc{Professor} \cite{Buckley:2009bj} 
tuning system, via analyses that are encoded in
\tsc{Rivet}~\cite{Buckley:2010ar}, for all shower models. This
relatively agnostic (automated) tuning 
approach also makes it possible to make (relatively) objective
statements concerning whether each shower model is able to describe
the existing data with a similar quality\footnote{With the caveats that
  not all relevant model parameters were included in the tuning, that the
  importance (weight) associated with (each bin of) each histogram is
  subjective, that the MC statistics are limited, and that the
  $\chi^2$ measure of difference does not take theoretical
  uncertainties of any kind into account.}.

The goodness-of-fit per degree of freedom provides information about how well
data measurements are described by the predictions of Monte Carlo (MC)
event
generators. It is defined as
\begin{align}
\frac{\chi^2}{N_{\text{dof}}}=
\frac{\sum\limits_\mathcal O w_\mathcal O\sum\limits_{b\in \mathcal O}(f_b(\vec p)-\mathcal R_b)^2/\Delta_b^2}
{\sum\limits_\mathcal O w_\mathcal O|{b\in\mathcal O}|}~,
\end{align} 
with reference value $\mathcal R_b$ and total error $\Delta_b$ of the data per bin $b$ 
and observable $\mathcal O$. The true MC response is modelled by a set of 
functions $f_{b}(\vec p)$. These functions are replaced by the true MC
response $\text{MC}_{b}(\vec p)$, if real MC runs are used.
The observables' weights $w_\mathcal O$ enter in the 
calculation of the goodness-of-fit as well as in the number of degrees of
freedom.

\subsection{Observables and Parameters} 

As observables for the tuning we use event shapes, identified-particle spectra, 
jet rates, particle multiplicities and $b$-quark fragmentation functions, 
provided by the \tsc{ALEPH}~\cite{Barate:1996fi,Heister:2001jg},
\tsc{DELPHI}~\cite{Abreu:1996na} and \tsc{OPAL}~\cite{Pfeifenschneider:1999rz} 
experiments and by the Particle Data Group \tsc{PDG}~\cite{Amsler:2008zzb}. 
The observables and their weights can be found in 
\TabsRef{tab:TuneObs_ES}-\ref{tab:TuneObs_B} in the appendix. 

The parameters for the hadronization and shower models of \tsc{Herwig++}, 
\tsc{Pythia~8} and \tsc{Vincia}, that we readjust here, 
can be found in \TabsRef{tab:TuneParamsH++} 
and \ref{tab:TuneParamsPV} in the appendix, together with a short
description.  

After performing a first tune with \tsc{Herwig++} we obtain flat
distributions in  $\chi^2$ for two parameters, the soft scale
$\mu_{\text{soft},FF}$ and the smearing parameter $\text{Cl}_\text{smr}$.
Therefore, we keep $\text{Cl}_\text{smr}$ fixed at its default value and set
$\mu_{\text{soft},FF}$ to zero for a slight increase of the value of
the shower cutoff.  
This approach leads to slightly smaller values in the goodness-of-fit values 
since the minimization works better due to the reduction of the
dimensionality of the parameter space.  

To get a good description of the MC response by the interpolation 
function of \tsc{Professor}, we use a fourth-order polynomial. Due to
fixing those parameters which exhibit flat distributions in $\chi^2$, as
explained above, we remain with six parameters for each combination  
of shower and hadronization model.  
The minimal possible number of MC runs needed for the tuning is defined
by the number of coefficients for the polynomial; here we need at least $210$
runs. To get reasonable results we perform oversampling 
of about a factor $3$, leading to $650$ MC runs with different randomly selected 
values of the parameters that are tuned. 
We use $500$ randomly selected runs $300$ times to interpolate the generator response
and check the quality of the interpolation by comparing the $\chi^2$ of the interpolation 
response with real MC runs at certain parameter values. 
By removing parameter regions where the interpolation did not work
sufficiently well we increase the quality of the interpolation. 
Unfortunately we cannot remove all bad regions for \tsc{Herwig++} 
since the values of some observables 
are not a smooth function of the gluon mass in the region 
where the MC predictions fit the data well. 
This is backed by the possibility of new splitting processes for higher gluon masses.
We use the $300$ different run combinations again in the tuning step 
where the goodness-of-fit is minimized in order to obtain the 
parameters that describe the observables best. Afterwards we perform real 
MC runs for these different parameter sets and calculate 
the real $\chi^2/N_\text{dof}$ to get the best tune.

\subsection{Tuning Results}

This section presents the results of the tuning process, starting with a short overview in terms 
of the total $\chi^2/N_\text{dof}$ values for the different shower
models. In order to validate  
the results of the tuning, we apply different analysis tools. The results 
for the $p^2_{\perp\mrm{dip}}$-ordered dipole shower are presented as
an example for \tsc{Herwig++}  
and for the $p_{\perp\mrm{ant}}^2$-ordered shower as an example for \tsc{Vincia}.
The parameter values obtained by the best tune are listed in the appendix, in 
\TabRef{tab:TuneResultsH++} for \tsc{Herwig++} and in \TabRef{tab:TuneResultsPV}
for \tsc{Pythia~8} and \tsc{Vincia}. In addition, the default values
and the scanned range are shown  
for the different parameters.

\subsubsection*{Quality of the Overall Description}
The goodness-of-fit function per degree of freedom,
$\chi^2/N_\text{dof}$, is listed in 
\TabRef{tab:OverallChi2} for each of the  shower models included in
the study, before and after tuning. The previous (default) 
tunes of \tsc{Vincia} and \tsc{Pythia~8} already describe the existing
LEP measurements very well. 
The description of the LEP data by the default angular-ordered tune of
\tsc{Herwig++} is fine as well. Therefore only small improvements in
the quality of the description of  
LEP data are achieved. Note that the angular-ordered shower 
is the only one that describes the mean particle multiplicities better
than the other observables. In the context of the string-based models,
one would presumably need to include the spin- and flavour-sensitive parameters
in the tuning as well, to reoptimize the agreement with the mean
identified-particle multiplicities. We did not look into this 
here, since the four-jet observables we investigate are not sensitive to
the particle composition, and since including these parameters would
have greatly inflated the dimensionality of the parameter space.

For the \tsc{Herwig++} dipole shower, for ordering in transverse momentum 
as well as for ordering in virtuality, the tuning greatly improved 
the quality of the description of 
the LEP data. The goodness-of-fit values are reduced by factors up to $17$. 

In terms of the overall description of the LEP data, \tsc{Vincia} with
ordering in transverse momentum fits the data the best,  
followed by \tsc{Pythia~8} and \tsc{Vincia} with
$m_{\mrm{ant}}^2$-ordering. Especially the two $p_\perp$-ordered
models achieve very similar $\chi^2/N_\mrm{dof}$ values and hence
cannot be told apart using the present data, nor does the 
mass-ordered version of \tsc{Vincia} stand out very
clearly after retuning. (Among the event shapes, the in- and
out-of-plane $p_\bot$ distributions exhibit the most significant
individual discrepancies with the data. We suspect colour-reconnection
effects may  
play a role for these distributions, an issue which is still very
actively investigated~\cite{Sjostrand:1993hi,Rathsman:1998tp,Skands:2007zg,Platzer:2012np,Gieseke:2012ft,Nagy:2012bt}.)
The three shower models interfaced to the cluster
hadronization model in \tsc{Herwig++} come in at
somewhat higher overall $\chi^2/N_\mrm{dof}$ values. 

We note that all the LEP measurements used 
\tsc{Pythia}~\cite{Sjostrand:1993yb} or
\tsc{Jetset}~\cite{Sjostrand:1993yb} to generate MC event samples for
the detector correction, hence there may be a small systematic bias
favouring the string-based models (here \tsc{Pythia~8} and
\tsc{Vincia}). 
\tsc{Herwig} event samples were used as well, to estimate the
systematic uncertainties. Therefore, the experiments claim that the
observable distributions are independent of the underlying MC
generator for the detector corrections within the experimental
systematics. 

\begin{table}[t]
\begin{center}
\begin{tabular}{lrr}
\toprule
 & \multicolumn{2}{l}{$\chi^2/N_{\text{dof}}$ for} \\ 
 & Default Parameter Values & Best Tune \\
\midrule
\tsc{Herwig++} $\tilde q^2$-Ordered Shower & 20.2 & 16.9  \\
\tsc{Herwig++} $p^2_{\perp\mrm{dip}}$-Ordered Dipole Shower & 348.5 & 23.0 \\
\tsc{Herwig++} $q^2_{\mrm{dip}}$-Ordered Dipole Shower & 358.2 & 25.3 \\
\tsc{Vincia} $p_{\perp\mrm{ant}}^2$-Ordered Shower & 7.1 & 6.4 \\
\tsc{Vincia} $m_{\mrm{ant}}^2$-Ordered Shower & 15.6 & 9.2 \\
\tsc{Pythia~8} $p^2_{\perp\mrm{evol}}$-Ordered Shower & 8.0 & 7.4 \\
\bottomrule
\end{tabular}
\end{center}
\caption{The total $\chi^2/N_\text{dof}$ values for the different shower models, 
for the default values of the parameters and the best tune.}
\label{tab:OverallChi2}
\end{table}

\subsubsection*{Validation}
The distribution of the $\chi^2/N_\text{dof}$ values of the $300$ tunes, each based 
on $500$ randomly selected runs at different parameter points,
are plotted in \FigsRef{fig:ScatterHPt} and \ref{fig:ScatterVPt} for two parameters for the 
\tsc{Herwig++} $p^2_{\perp\mrm{dip}}$-ordered dipole shower and for \tsc{Vincia} with ordering 
in $p_{\perp\mrm{ant}}^2$.
Narrow distributions indicate that the observables are very sensitive to this parameter.
Broader distributions are obtained if either the observables are less sensitive to 
a parameter or, as for the Lund parameters $a_L$ and $b_L$, if two parameters 
are highly correlated. 

\begin{figure}[p]
\centering
\input{images/Tuning/AlphaMZ_scatter.tex} 
\input{images/Tuning/PSplit_scatter.tex}
\caption{Scatterplots for \textsf{AlphaMZ} and \textsf{PSplit} with
  real MC runs for the  
\tsc{Herwig++} $p^2_{\perp\mrm{dip}}$-ordered dipole shower. The plots show the $\chi^2/N_\text{dof}$
values of the $300$ different run combinations with respect to the parameter value.
The vertical line indicates the parameter value of the best tune and the plot boundaries are chosen to be equal to
the scanned range of the parameter.}
\label{fig:ScatterHPt}\vspace*{3mm}
\input{images/Tuning/aLund_scatter.tex} 
\input{images/Tuning/PTsigma_scatter.tex}
\caption{Scatterplots for \textsf{aLund} and \textsf{PTsigma} with real MC runs for the 
\tsc{Vincia} $p_{\perp\mrm{ant}}^2$-ordered shower. The plots show the $\chi^2/N_\text{dof}$
values of the $300$ different run combinations with respect to the parameter value.
The vertical line indicates the parameter value of the best tune and the plot boundaries are chosen to be equal to
the scanned range of the parameter.}
\label{fig:ScatterVPt}
\end{figure}

\begin{figure}[p]
\centering
\input{images/Tuning/AlphaMZ.tex} 
\input{images/Tuning/PSplit.tex}
\caption{A scan of \textsf{AlphaMZ} and \textsf{PSplit} with real MC runs and the interpolation result of  
\tsc{Professor} for the \tsc{Herwig++} $p^2_{\perp\mrm{dip}}$-ordered dipole shower. All other parameters
are fixed at their new tuned values. The vertical line indicates the value of the 
best tune of the scanned parameter. The curves show the $\chi^2/N_\text{dof}$ 
for the different types of observables and the blue curve the combination of all observables. 
Points correspond to the real MC and lines to the interpolation result.}
\label{fig:ScanChi2HPt}\vspace*{3mm}
\input{images/Tuning/aLund.tex} 
\input{images/Tuning/PTsigma.tex}
\caption{A scan of \textsf{aLund} and \textsf{PTsigma} with real MC runs and the interpolation result of  
\tsc{Professor} for the \tsc{Vincia} $p_{\perp\mrm{ant}}^2$-ordered shower. All other parameters 
are fixed at their new tuned values. The vertical line indicates the value of the 
best tune of the scanned parameter. The curves show the $\chi^2/N_\text{dof}$ 
for the different types of observables and the blue curve the combination of all observables. 
Points correspond to the real MC and lines to the interpolation result.}
\label{fig:ScanChi2VPt}
\end{figure}

In order to verify the result of the generator tuning with \tsc{Professor} 
we perform real MC runs where we change only one parameter with 
randomly distributed values and set all other parameters to their 
new tuned value. We reproduce the histograms at the same parameter points 
by using the interpolation function calculated by \tsc{Professor} 
to model the MC response.
The distribution of the goodness-of-fit is shown with respect to the parameter
value for two different parameters for the \tsc{Herwig++} 
$p^2_{\perp\mrm{dip}}$-ordered dipole shower and for \tsc{Vincia} with ordering 
in $p_{\perp\mrm{ant}}^2$ in \FigsRef{fig:ScanChi2HPt} and \ref{fig:ScanChi2VPt}. 
The $\chi^2/N_\text{dof}$ value is split for the 
different groups of observables where the lines correspond 
to the interpolation result and the points to the real MC runs. 
The single observables enter in the calculation of the goodness-of-fit 
for a group of observables with the same weight as for the calculation of the 
overall $\chi^2/N_\text{dof}$.
By comparing the interpolation with the real generator response, the 
quality of the interpolation function can be evaluated as well.
\FigsRef{fig:ScanChi2HPt} and \ref{fig:ScanChi2VPt} show that the 
parameter values of the best tune, marked by the vertical line, are clearly 
favoured, mostly driven by event shapes. As mentioned above, we were not 
able to remove all regions for \tsc{Herwig++} where the interpolation
did not work sufficiently well. This leads to the different $\chi^2/N_\text{dof}$ values for 
interpolation and MC runs. Since the quality of the interpolation is disrupted 
by the possibility for new splitting processes for higher gluon masses,  
identified particle spectra and mean multiplicities 
cannot be described very well. This affects of course also the other parameters.
As shown in \FigRef{fig:ScanChi2VPt}, the interpolation works better 
for \tsc{Vincia}, where interpolation and generator response agree perfectly. 

Besides the $\chi^2/N_\text{dof}$ distribution of the parameters we 
have shown here, we obtain parameters with flatter distributions as well.
In addition some parameters prefer to be at the 
limit of the scanned range as occurring for example 
for the strong coupling $\alpha_{S}$ within the tuning of \tsc{Pythia~8} and \tsc{Vincia}
with $m_{\mrm{ant}}^2$-ordering.

\subsection{Eigentunes}

To estimate the uncertainty in the MC predictions in connection with
changing the parameter values during the tuning, so-called eigentunes are performed.
The parameters are varied along the eigenvectors in parameter space where the eigenvectors 
are obtained by certain changes, $\Delta \chi^2/N_\text{dof}$, in $\chi^2/N_\text{dof}$. 
For each parameter two eigenvectors, one in the ``$+$'' and one in the ``$-$'' direction, exist. 
If the goodness-of-fit were distributed as a true $\chi^2$ function 
$\Delta\chi^2/N_\text{dof}=1$ would correspond to a one sigma deviation and 
$\Delta\chi^2/N_\text{dof}=4$ to a two sigma deviation from the
minimum (i.e., the central tune), etc. 

Given, however, that none of the models achieves a $\chi^2/N_\text{dof}
\le 1$, the eigentunes can at most be used to give a rough indication
of the range of accessible model variations near the respective
minimum for each model. This is still valuable, as it can help us 
determine whether the central tunes of two (or more) different theory
models could easily be retuned to give the same result or not, on a given
observable (overlapping versus non-overlapping eigentune variation
ranges). 

We calculate two sets of eigentunes with \tsc{Professor},
corresponding to one- and two-sigma deviations, and perform MC runs to obtain
envelopes around the central tune for each of the six different theory models.
For the four-jet observables we
propose (see \secRef{sec:results}), we find that the model differences are \emph{larger} than the
individual eigentune
envelopes.  Hence we conclude that these observables 
\emph{do} have sensitivity to distinguish
between the theory models within the limits of the tuning uncertainty.
For all further studies we will use the central tunes.

\section{Results \label{sec:results}}
We consider hadronic $Z$ events (photon ISR is switched off)
and use the Durham $k_T$ clustering algorithm~\cite{Catani:1991hj} to
cluster all events back to two jets, keeping track of the intermediate
clustering scales along the way. 
The $3\to 2$ clustering scale is
denoted $y_{23} = k_{T3}^2/m_Z^2$, and so on for 
higher jet numbers. We require both $y_{23}$ and $y_{34}$ to be
greater than $0.0045$, to obtain an inclusive 4-jet event sample with minimal 
contamination from $B$ decays and lower (non-perturbative) scales.

Strong ordering corresponds to $y_{23}\gg y_{34} \gg \ldots$, while
events with, e.g., $y_{34}\sim y_{23}$  should be more
sensitive to the ordering condition and to the effective $1\to 3$
splitting kernels. 
Further, we may in principle also keep track of which ``side'' each
clustering step happens on, which can give us an additional handle on
the relative contributions to 
the four-jet rate from ``opposite-side'' $1\to 2 \otimes 1\to 2$
splittings versus ``same-side'' $1\to 3$ ones. Within the context of
this study, however, we only explicitly used the former requirement
($y_{23}\sim y_{34}$), in the context of the definition of the
$M_L^2/M_H^2$ variable, though we note that 
the latter (same-side vs opposite-side sequential clusterings) is
implicitly present along the $M_L^2/M_H^2$ axis. 

\subsection{Observable 1: $\mathbf{\theta_{14}}$ \label{sec:obs1}}

\begin{figure}[t]
\centering
\includegraphics[width=7.5cm]{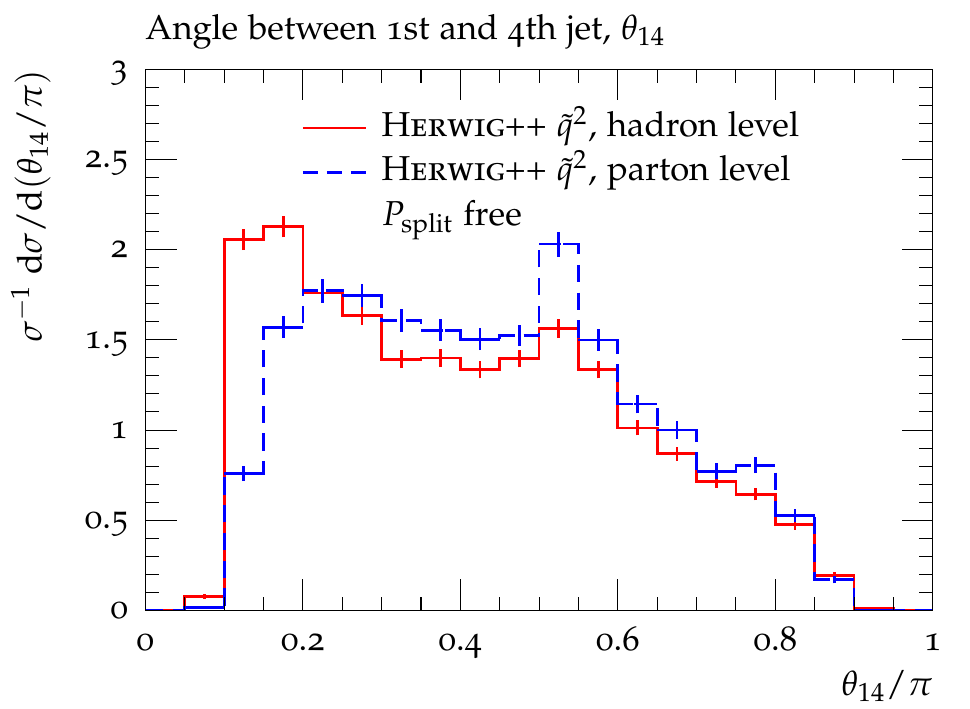}\hspace*{2mm}
\includegraphics[width=7.5cm]{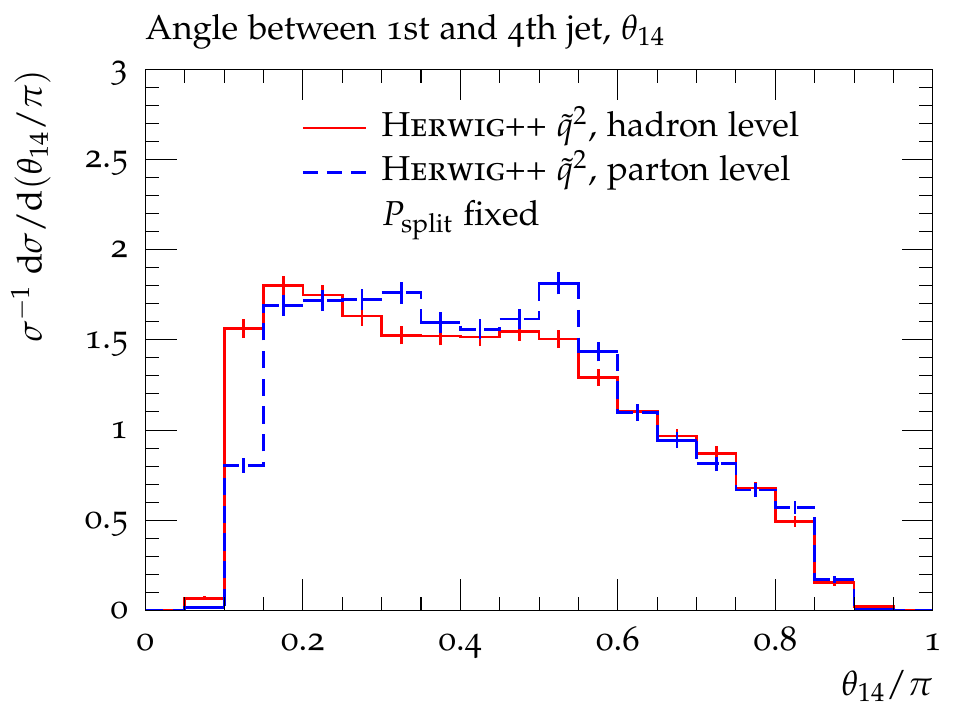} \\
\includegraphics[width=7.5cm]{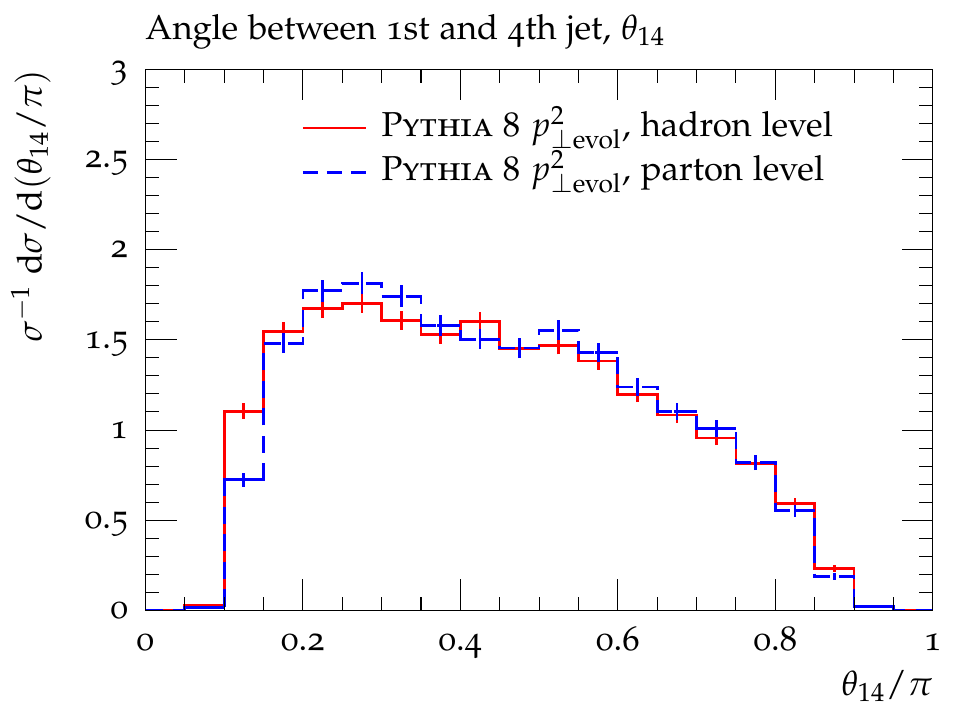}\hspace*{2mm}
\includegraphics[width=7.5cm]{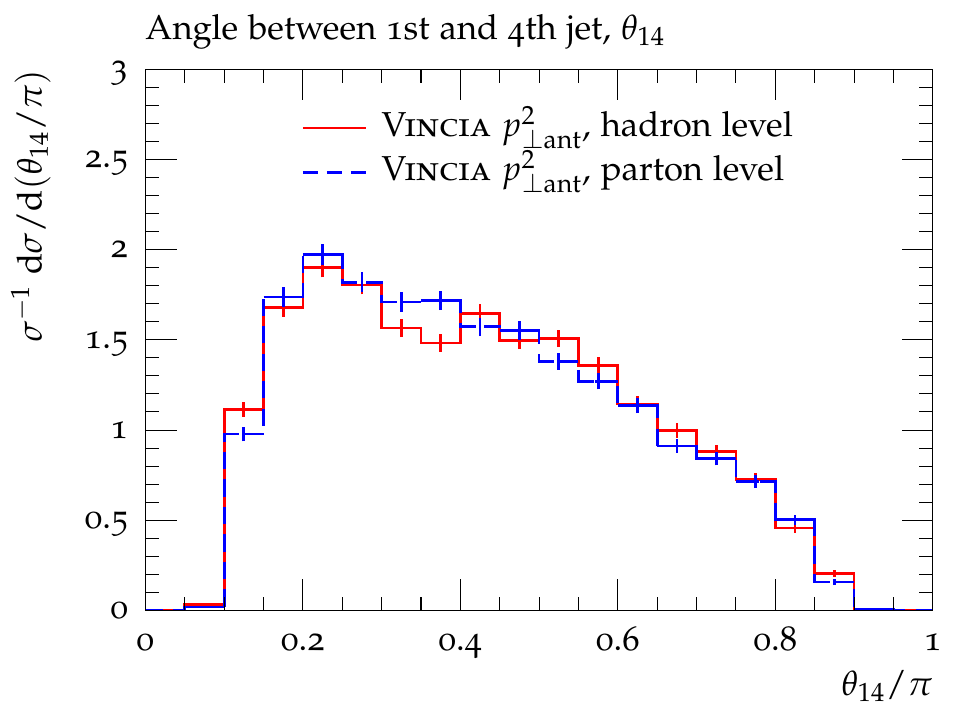}
\caption{A comparison of the normalized distribution of $\theta_{14}$ on hadron 
(red solid) and parton level (blue dashed).
The upper row shows the predictions of the \tsc{Herwig++} angular ordered shower,
where $P_\text{split}$ is kept fixed in the right plot and 
no constraints in the left plot.
In the lower row the same plot is shown for \tsc{Pythia~8} and the 
$p_{\perp\mrm{ant}}^2$-ordered shower of \tsc{Vincia}.}
\label{fig:HadPartLevel}
\end{figure}

We consider the event at the stage when it has been clustered back to 
four jets, and order the jets in hardness. To be sensitive to coherence
we constrain the angles between the jets such that the first (hardest)
jet lies back-to-back to a near-collinear jet pair, formed by the
second and  
third jet; $\theta_{12}>2\pi/3$, $\theta_{13}>2\pi/3$ and $\theta_{23}<\pi/6$.
From this near-collinear three-jet state we probe the emission angle
of the soft fourth jet with respect to the first jet, $\theta_{14}$. 

Before presenting the main results for $\theta_{14}$, we note that, 
for \tsc{Herwig++} with default shower and hadronization parameters 
an enhancement for small values of $\theta_{14}$ shows up due to
surprisingly large  non-perturbative effects. This enhancement
decreases for the dipole shower due to  
changing the values of the hadronization parameters throughout the tuning, 
but unfortunately not for the angular-ordered shower. By checking the 
influence of the hadronization parameters on the shape of the distribution 
of $\theta_{14}$, we identify the mass exponent for daughter clusters, 
$P_\text{split}$, as the cause of the enhancement. By keeping it fixed 
at a value of $0.6$ during the tuning, we achieve a better agreement 
between hadron level and parton level for the normalized distribution 
of $\theta_{14}$, which we regard as physically more reasonable given
the cut of $y_{34}>0.0045$. This distribution is shown in
the upper row of \FigRef{fig:HadPartLevel} 
for keeping $P_\text{split}$ fixed on the right and for no constraints 
on the left. We see that the influence of hadronization is 
reduced strongly by keeping the hadronization parameter fixed.
However, some sensitivity to non-perturbative effects is still left
for small values of the observable $\theta_{14}$. The \tsc{Herwig++}
dipole shower gives similar results in the comparison of
the angular observables on hadron and parton level. 
In addition we show this comparison in the lower row of 
\FigRef{fig:HadPartLevel} for \tsc{Pythia~8} and the 
$p^2_{\perp\mrm{ant}}$-ordered shower of \tsc{Vincia}. 
For \tsc{Pythia~8} we can see a small enhancement for 
small values of $\theta_{14}$ as well, whereas the predictions of
\tsc{Vincia} agree well with each other.

To compare the predictions of the different theory models, \FigRef{fig:ResObs1}
shows the normalized distribution of $\theta_{14}$ in the upper left plot. 

\begin{figure}[tp]
\centering
\includegraphics[width=7.5cm]{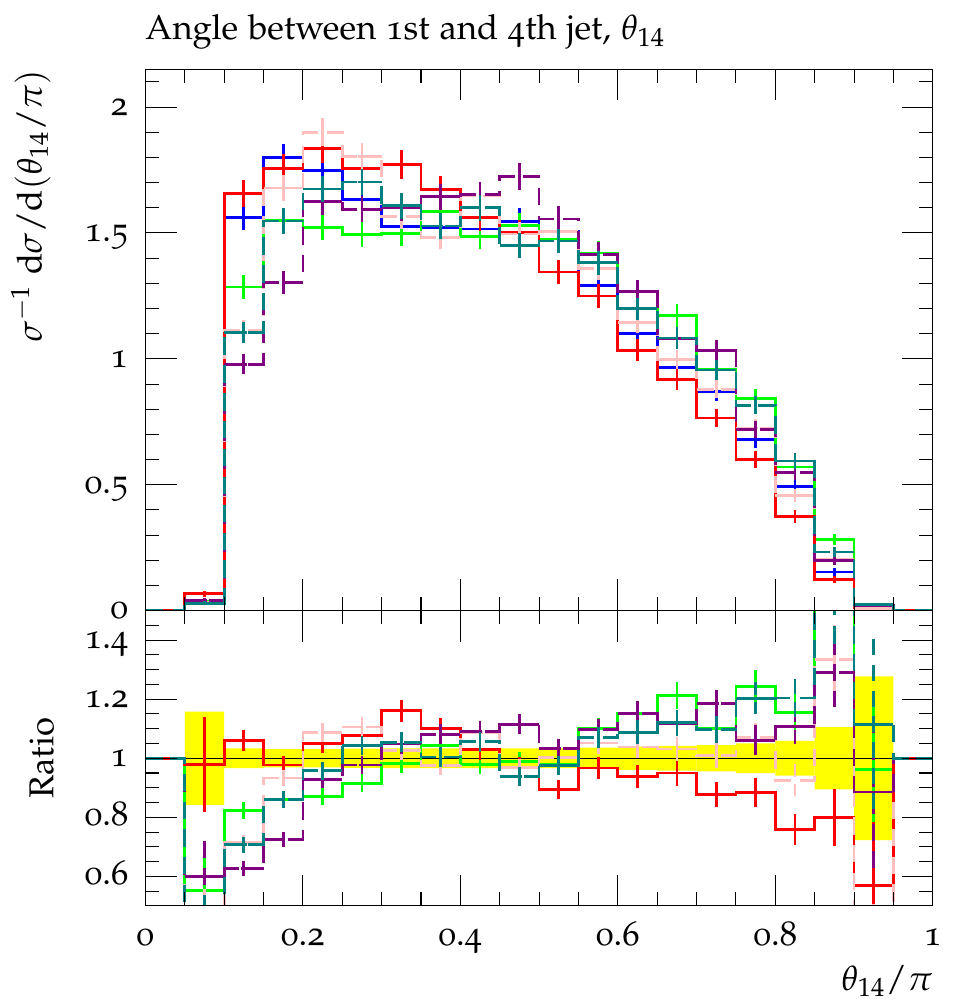}\hspace*{2mm}
\includegraphics[width=7.5cm]{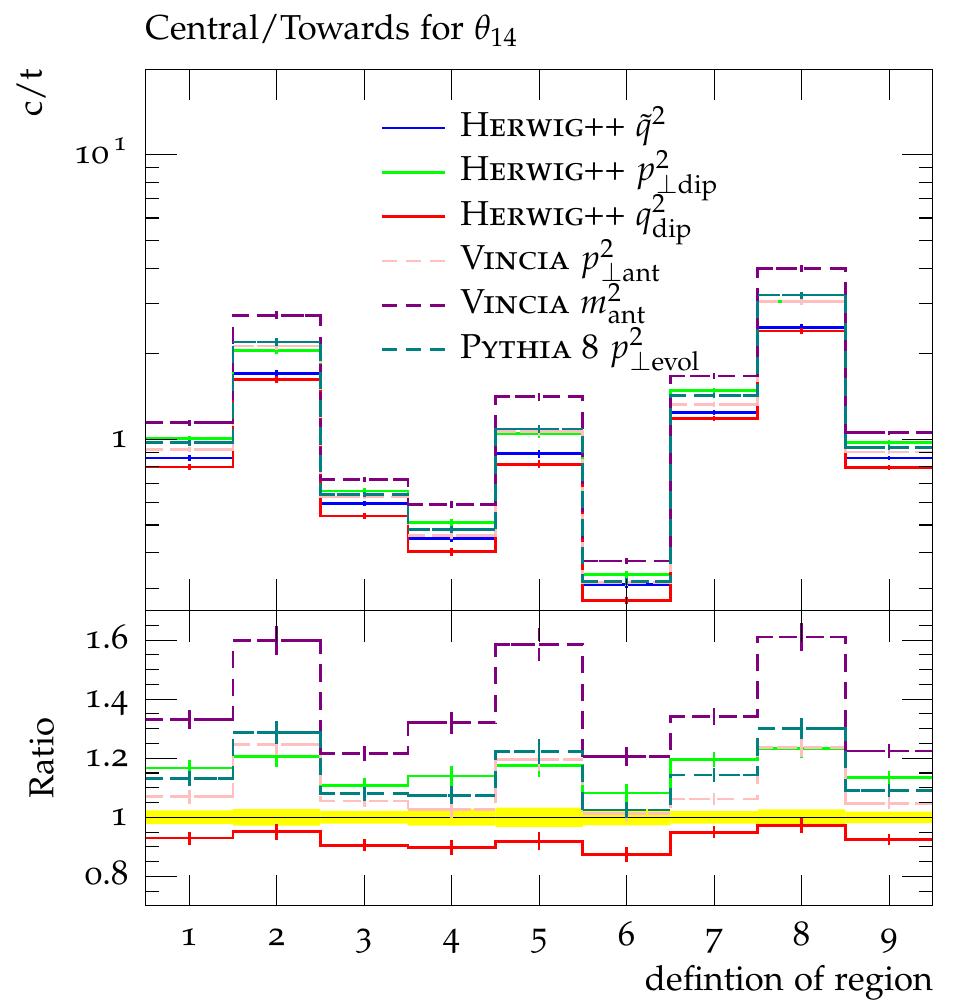} \\
\includegraphics[width=7.5cm]{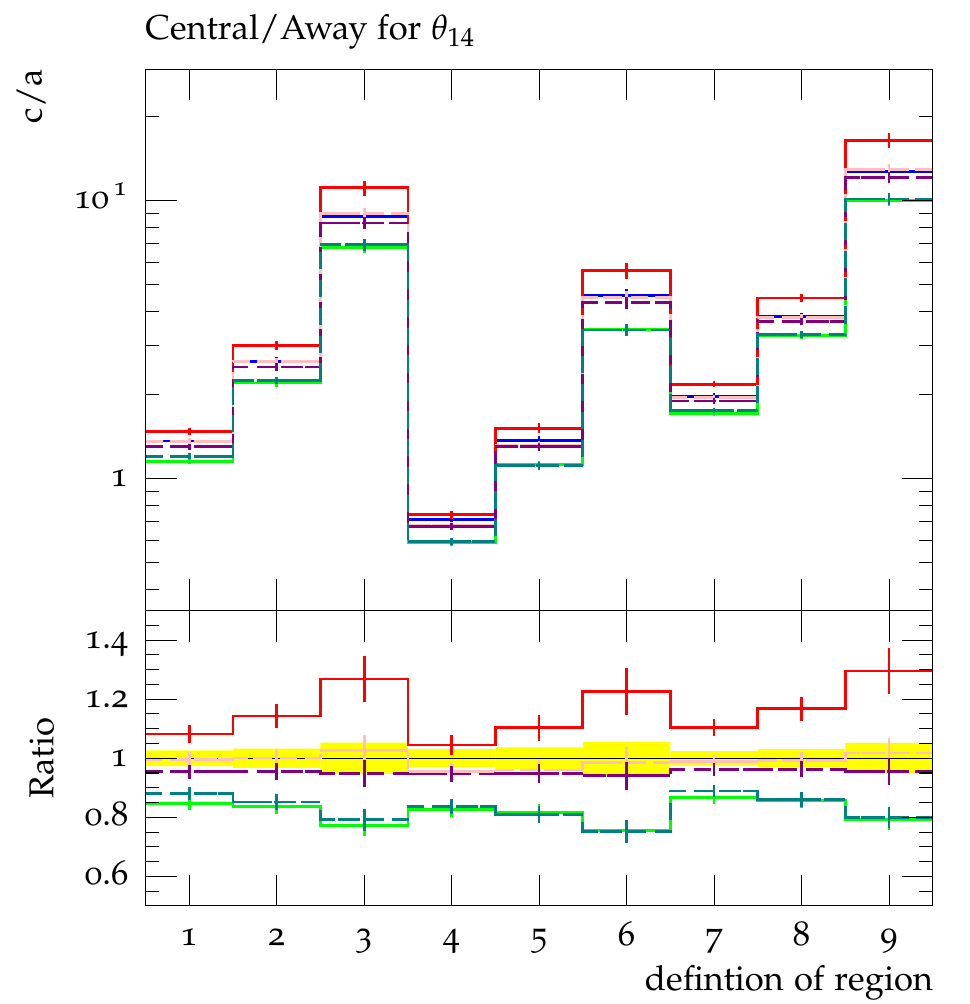}\hspace*{2mm}
\includegraphics[width=7.5cm]{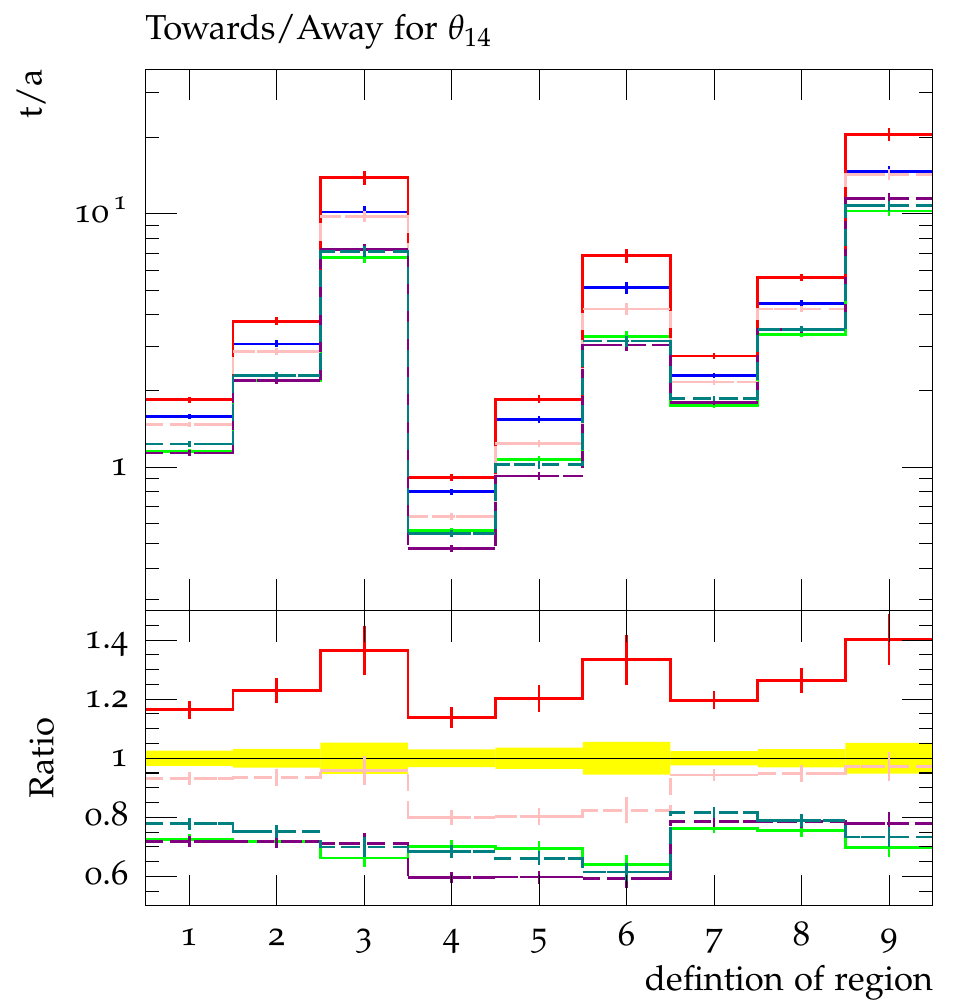}
\caption{{\sl Upper Left:} normalized distribution of the 
angular observable $\theta_{14}$. {\sl Other plots:} ratios of
different regions with respect to the definition of the region,
cf~\TabRef{tab:DefRegionsA14}. 
The solid curves refer to the \tsc{Herwig++} showers, the angular-ordered 
default shower in blue, the $p^2_{\perp\mrm{dip}}$-ordered in green
and the $q_\mrm{dip}^2$-ordered dipole shower in red
respectively. The dashed lines refer to the \tsc{Vincia}  
shower with $m_{\mrm{ant}}^2$-ordering in violet and
$p_{\perp\mrm{ant}}^2$-ordering in pink and to the \tsc{Pythia~8} shower in teal.
The ratio plots show the deviation of the showers with respect to the
\tsc{Herwig++} angular-ordered default shower.
The vertical error bars indicate the expected $1\sigma$ statistical error with 
$5\cdot10^5$ hadronic $Z$ decays.}
\label{fig:ResObs1}
\end{figure}

To show the differences more clearly and reduce the observable to a
simpler quantity with better statistics, we divide the full
$\theta_{14}$ range into three regions labelled ``Towards'' (small
$\theta_{14}$), ``Central'' (intermediate $\theta_{14}$), 
and ``Away'' (large $\theta_{14}$). We may then consider the ratio between
regions,  
\begin{align}
AS(x) = \dfrac{\sum\limits_{x_1<x<x_2}y(x)}{\sum\limits_{x_3<x<x_4}y(x)}~.
\end{align}
In the ``Towards'' region, the first and fourth jet are collinear,
while they are back-to-back in the ``Away'' region.
Events where the fourth jet is a wide-angle emission from the three-jet system
populate the ``Central'' region. We consider nine different possibilities
for the exact divisions between the regions, listed in
\TabRef{tab:DefRegionsA14} and corresponding roughly to looser or
tighter cuts. The ratios of the integrated $\theta_{14}$ rates 
for each of the nine different region definitions are shown in
\FigRef{fig:ResObs1}.  

\begin{table}[t]
\begin{center}
\begin{tabular}{ccccc} 
 & & Central/Towards & Central/Away & Towards/Away \vspace*{2mm} \\
\toprule
\# & Central Region & Towards Region & Away Region & Towards region \\ 
\midrule
1 & $0.4<\theta_{14}/\pi<0.6$ & $\theta_{14}/\pi<0.3$ & $\theta_{14}/\pi>0.6$ & $\theta_{14}/\pi<0.3$ \\
2 & $0.4<\theta_{14}/\pi<0.6$ & $\theta_{14}/\pi<0.2$ & $\theta_{14}/\pi>0.7$ & $\theta_{14}/\pi<0.3$ \\
3 & $0.4<\theta_{14}/\pi<0.6$ & $\theta_{14}/\pi<0.4$ & $\theta_{14}/\pi>0.8$ & $\theta_{14}/\pi<0.3$ \\
4 & $0.45<\theta_{14}/\pi<0.55$ & $\theta_{14}/\pi<0.3$ & $\theta_{14}/\pi>0.6$ & $\theta_{14}/\pi<0.2$ \\
5 & $0.45<\theta_{14}/\pi<0.55$ & $\theta_{14}/\pi<0.2$ & $\theta_{14}/\pi>0.7$ & $\theta_{14}/\pi<0.2$ \\
6 & $0.45<\theta_{14}/\pi<0.55$ & $\theta_{14}/\pi<0.4$ & $\theta_{14}/\pi>0.8$ & $\theta_{14}/\pi<0.2$ \\
7 & $0.35<\theta_{14}/\pi<0.65$ & $\theta_{14}/\pi<0.3$ & $\theta_{14}/\pi>0.6$ & $\theta_{14}/\pi<0.4$ \\
8 & $0.35<\theta_{14}/\pi<0.65$ & $\theta_{14}/\pi<0.2$ & $\theta_{14}/\pi>0.7$ & $\theta_{14}/\pi<0.4$ \\
9 & $0.35<\theta_{14}/\pi<0.65$ & $\theta_{14}/\pi<0.4$ & $\theta_{14}/\pi>0.8$ & $\theta_{14}/\pi<0.4$ \\
\bottomrule
\end{tabular}
\end{center}
\caption{Definition of the different regions for the asymmetry of
  $\theta_{14}$. Columns 2--5 specify the limits for the regions and
  the first column gives the numbering.  
The ratio of the central to towards region is built with the 2nd
and 3rd column, central to away with the 2nd and 4th and
towards to away uses the 4th and 5th column.} 
\label{tab:DefRegionsA14}
\end{table}

Since large non-perturbative effects occur in the towards region for the 
\tsc{Herwig++} shower models, we consider the ratio of the central to 
away region to be the most robust observable. This ratio reflects the
relative amount 
of soft wide-angle emissions to emissions where the first jet lies back-to-back
to all other jets in the event. Compared to the angular-ordered shower,
the \tsc{Herwig++} dipole shower with $q_\mrm{dip}^2$-ordering
predicts up to $30\%$ higher values for this ratio; a very significant
difference. The predictions of the
$p^2_{\perp\mrm{dip}}$-ordered  
dipole shower of \tsc{Herwig++} are very similar to the ones of \tsc{Pythia~8} 
and lower than the predictions of all other shower models. 
\tsc{Vincia} with both ordering variables agrees with the result
of the \tsc{Herwig++} angular-ordered shower within the statistical errors.

To distinguish between the two evolution variables of \tsc{Vincia} we can use
the ratio of the central to towards region. This ratio 
reflects the relative amount of soft wide-angle emission compared to
collinear emission. The predictions of the $m_{\mrm{ant}}^2$-ordered shower 
are about $35\%$ higher than the ones of the $p_{\perp\mrm{ant}}^2$-ordered 
shower. We expect this behaviour since the $m_{\mrm{ant}}^2$-ordered 
shower prefers wide-angle soft over collinear 
emissions, see \FigRef{fig:evolution}, 
whereas the $p_{\perp\mrm{ant}}^2$-ordered shower prefers the opposite.

The third ratio, shown in the lower right plot of \FigRef{fig:ResObs1},
is the towards over away region, which is predicted very similarly by 
\tsc{Pythia~8}, \tsc{Herwig++} with $p^2_{\perp\mrm{dip}}$-ordering and 
\tsc{Vincia} with $m_{\mrm{ant}}^2$-ordering. The predictions of these
theory models are $20\%$ to $30\%$ smaller than the one of the 
angular-ordered shower of \tsc{Herwig++}. The $q_\mrm{dip}^2$-ordered
shower on the other hand produces values up to $40\%$ higher.
In both ratios including the away region, we see a $10\%$ to $20\%$
difference between the predictions of \tsc{Pythia~8} and the 
$p^2_{\mrm{ant}}$-ordered shower of \tsc{Vincia}, where \tsc{Pythia~8}
produces more events populating the away region. Thus, we conclude
that these ratios have significant discriminating power between the
models, including between \tsc{Pythia~8} and \tsc{Vincia} which 
appeared very similar in the global analysis. 

\subsection{Observable 2: $\mathbf{\theta^*}$}

\begin{figure}[t]
\centering
\includegraphics[width=7.5cm]{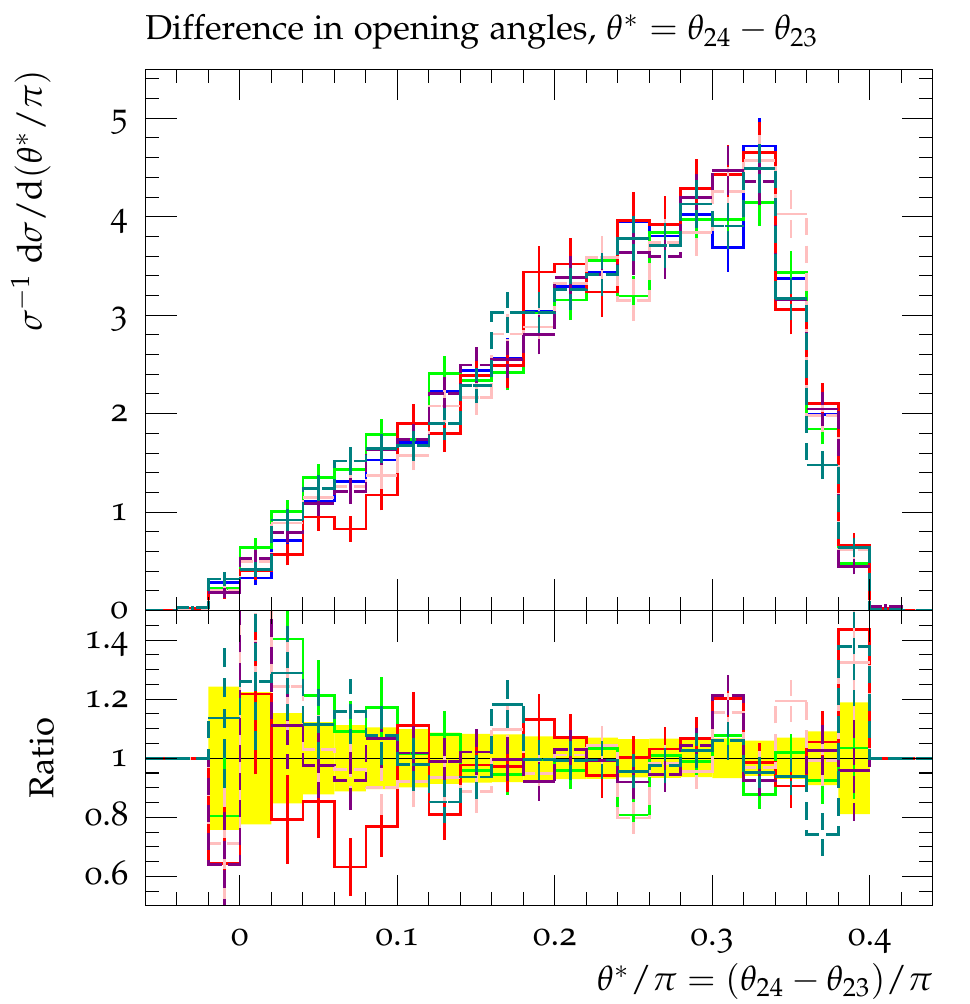}\hspace*{2mm}
\includegraphics[width=7.5cm]{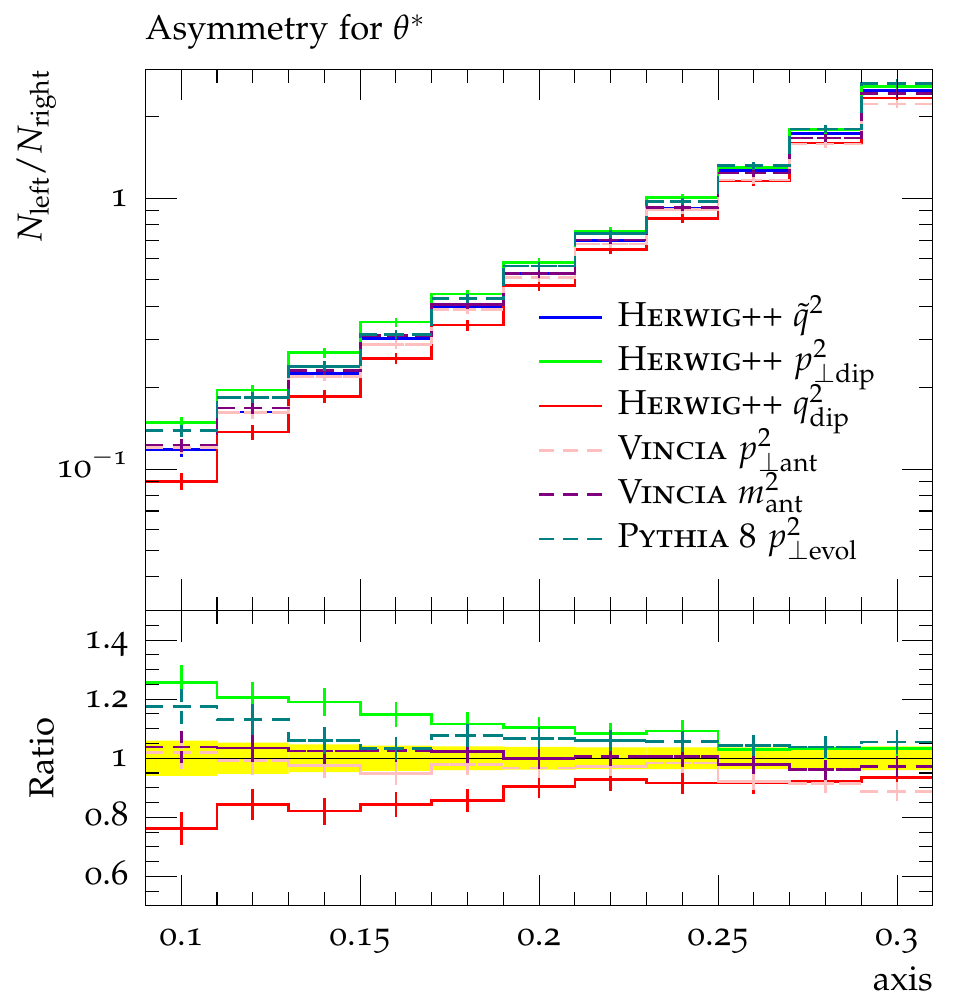}
\caption{The plots show the normalized distribution of the difference in opening angles, 
$\theta^*$, on the left and its asymmetry with respect to the
  asymmetry axis, $x_0$, on the right.}
\label{fig:ResObs2}
\end{figure}

In addition to the cuts for the previous observable, we require the fourth 
jet to be close in angle to the near-collinear (23) jet pair,
$\theta_{24}<\pi/2$, in order to enhance the sensitivity to coherent
emission off the (23) jet system. We then define our second angular observable 
as the difference in opening angles, 
$\theta^*=\theta_{24}-\theta_{23}$, and, similarly to above, 
we introduce the asymmetry,
\begin{align}\label{eq:asymmetry}
\frac{N_\text{left}}{N_\text{right}} = \dfrac{\sum\limits_{x<x_0}y(x)}{\sum\limits_{x>x_0}y(x)} ~,
\end{align}
with respect to an arbitrary dividing point, $\theta^*=x_0$, which
separates the  small-$\theta^*$ region from the large-$\theta^*$ one.
The normalized distribution of $\theta^*$ and the asymmetry (as a
function of the dividing point $x_0$) are 
shown in \FigRef{fig:ResObs2}. Due to the additional cut on
$\theta_{24}$ for this observable, the error bars are higher and thus
the statistical power in discriminating 
the different theory models smaller. The only shower model which can
be distinguished from the others is the $q_\mrm{dip}^2$-ordered
dipole shower of \tsc{Herwig++}. 
This model tends to predict more events where the difference in opening angles 
of the fourth and third jet is large, compared to the angular-order shower.
The $p^2_{\perp\mrm{dip}}$-ordered dipole shower of \tsc{Herwig++} predicts
larger values for the asymmetry and therefore more events with a smaller
difference in opening angles.

\subsection{Observable 3: $\mathbf{C_2^{(1/5)}}$}

\begin{figure}[t]
\centering
\includegraphics[width=7.5cm]{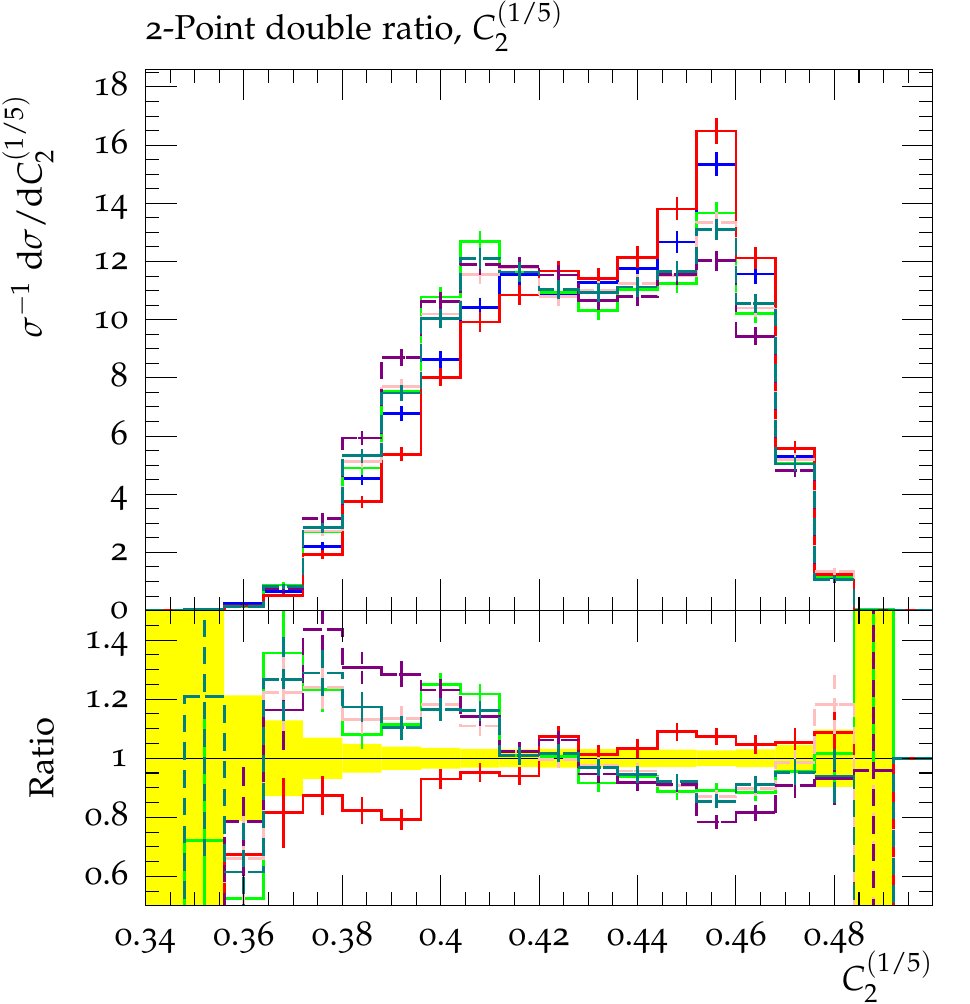}\hspace*{2mm}
\includegraphics[width=7.5cm]{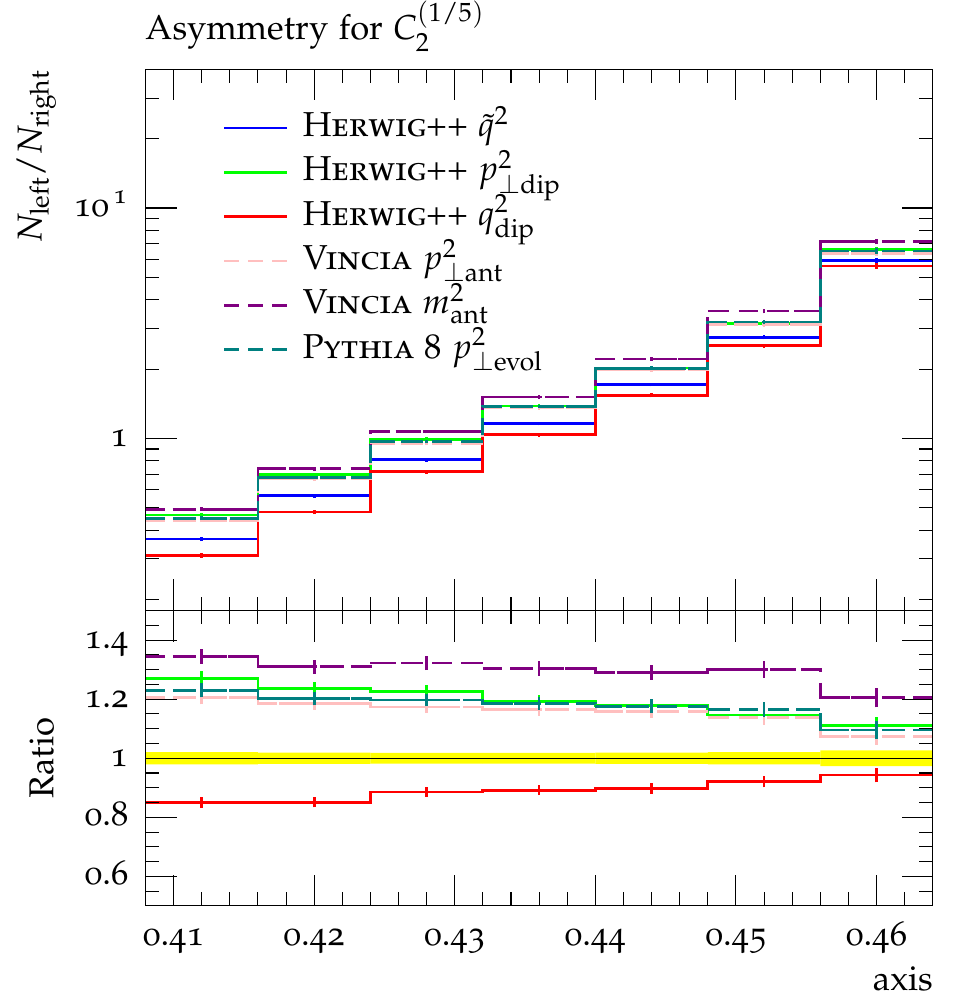}
\caption{The plots show the normalized distribution of the 2-point double ratio, 
$C_2^{(1/5)}$, on the left and its asymmetry with respect to the asymmetry axis, $x_0$,
 on the right.}
\label{fig:ResObs4}
\end{figure}

Ref.~\cite{Larkoski:2013eya} defines the $N$-point energy correlation 
function (ECF) as
\begin{align}
\text{ECF}(N,\beta)=\sum_{i_1<i_2<...<i_N}\left(\prod_{a=1}^NE_{i_a}\right)\left(\prod_{b=1}^{N-1}\prod_{c=b+1}^N
\theta_{i_bi_c}\right)^\beta~. \label{eq:ECF}
\end{align}
where the sum runs over all particles of a jet. To be sensitive
to the global event structure we replace this sum by the sum 
over all jets in the event. Thus, $\theta_{i_1i_2}$ 
denotes the angle between two jets $i_1$ and $i_2$.
The ECFs are used to build double ratios
\begin{align}
C_N^{(\beta)}= \frac{\text{ECF}(N+1,\beta)\text{ECF}(N-1,\beta)}{\left(\text{ECF}(N,\beta)\right)^2}~.
\end{align}
We choose a value of $\beta=1/5$ to give all angles about equal weights
and to be sensitive to soft configurations. Sensitivity to 
collinear configurations can be achieved by choosing $\beta=2$ and 
giving greater angles more weight.

In the 4-jet events described in \secRef{sec:obs1}
we use the 2-point double ratio
\begin{align} \label{eq:C2beta}
C_2^{(\beta)}=\frac{\sum\limits_{j_1<j_2<j_3}E_{j_1}E_{j_2}E_{j_3}(\theta_{j_1j_2}\theta_{j_1j_3}\theta_{j_2j_3})^\beta}
{\Big(\sum\limits_{j_1<j_2}E_{j_1}E_{j_2}\theta_{j_1j_2}^\beta\Big)^2}\cdot E_\text{vis}~,
\end{align}
where the sums run over the four jets. Due to the cuts on the angles
between the jets, the events look like three-jet systems to the observable.
This system contains two hard jets, jet $1$ and jet $(23)$\footnote{The 
combination of the second and third jet.}, lying approximately back-to-back
and a third soft jet, jet $4$. With this notation \EqRef{eq:C2beta} 
can approximately be written as
\begin{align} \label{eq:C2betaApprox1}
C_2^{(\beta)} \approx \frac{E_1E_{23}E_4(\theta_{1\,23}\theta_{14}\theta_{23\,4})^\beta}
{(E_1E_{23}\theta_{1\,23}^\beta+E_1E_4\theta_{14}^\beta+E_{23}E_4\theta_{23\,4}^\beta)^2}\cdot E_\text{vis}~.
\end{align}
Taking the small energy of jet $4$ and the large angle $\theta_{1\,23}>2\pi/3$
into account, the denominator can be reduced to its first term,
\begin{align}\label{eq:C2betaApprox2}
C_2^{(\beta)} \approx \frac{E_4(\theta_{14}\theta_{23\,4})^\beta}{E_1E_{23}\theta_{1\,23}^\beta}\cdot E_\text{vis}~.
\end{align}
This leaves only the angles relative to the fourth jet and the 
energies as free parameters.
For $\beta=1/5$ all angles are weighted relatively equal and 
hence $C_2^{(1/5)}$ is proportional to the energy of the fourth jet, 
relative to the remaining energy of the event. 

For the normalized distribution of the 2-point double ratio, 
$C_2^{(1/5)}$, we see non-perturbative effects for \tsc{Herwig++}
and the $m_{\mrm{ant}}^2$-ordered shower of \tsc{Vincia}.
For all of these shower models, hadronization and decays enlarge
the number of events with a harder fourth jet, hence higher values
of $C_2^{(1/5)}$. 

We again use the 
asymmetry, as defined in \EqRef{eq:asymmetry}, to condense 
the differences  between the theory models into a ratio of integrals. 
The normalized distribution of the 2-point double ratio, 
$C_2^{(1/5)}$, and the according asymmetry are shown in \FigRef{fig:ResObs4}.
As indicated by the asymmetry, the $p^2_\perp$-ordered shower models of 
\tsc{Herwig++}, \tsc{Pythia~8} and \tsc{Vincia} give similar
predictions. Compared to that, the prediction of the $m_{\mrm{ant}}^2$-ordered
shower of \tsc{Vincia} is higher, as expected due to the preference of
soft over collinear emissions during the population of phase-space.

\subsection{Observable 4: $\mathbf{M_L^2/M_H^2}$}

\begin{figure}[t]
\centering
\includegraphics[width=7.5cm]{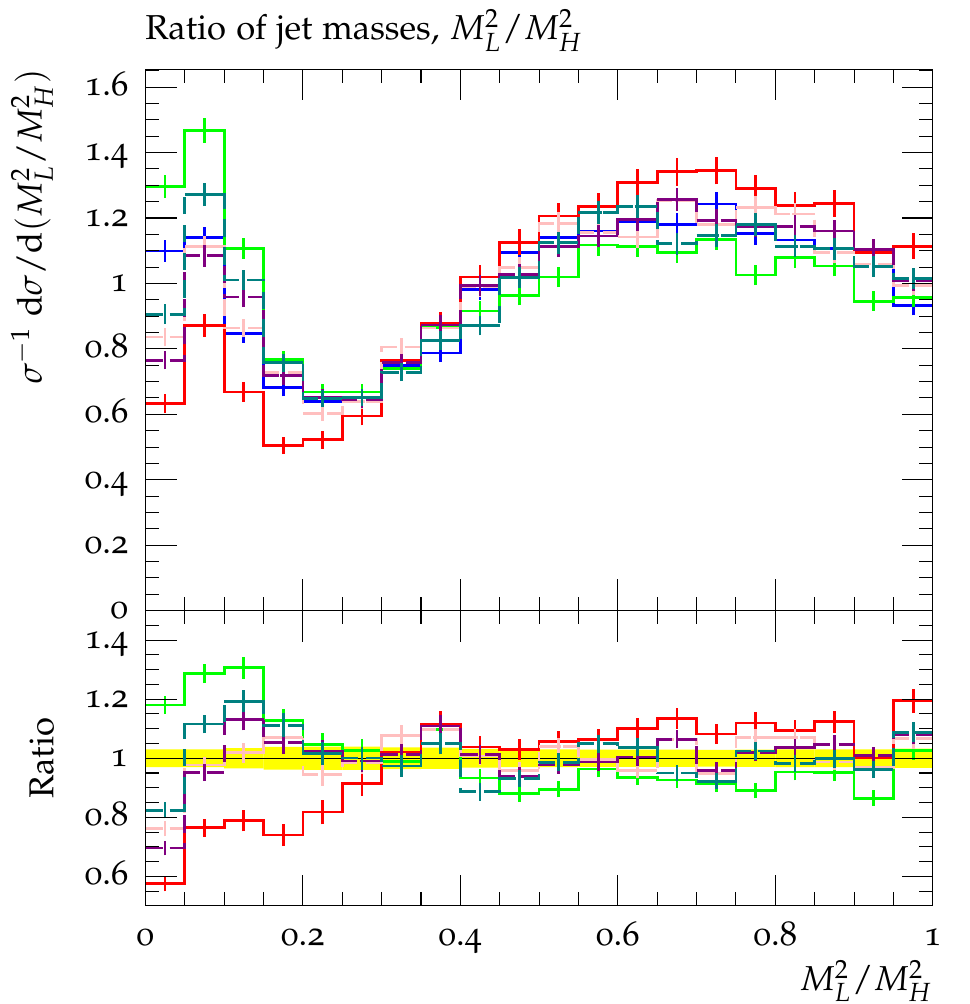}\hspace*{2mm}
\includegraphics[width=7.5cm]{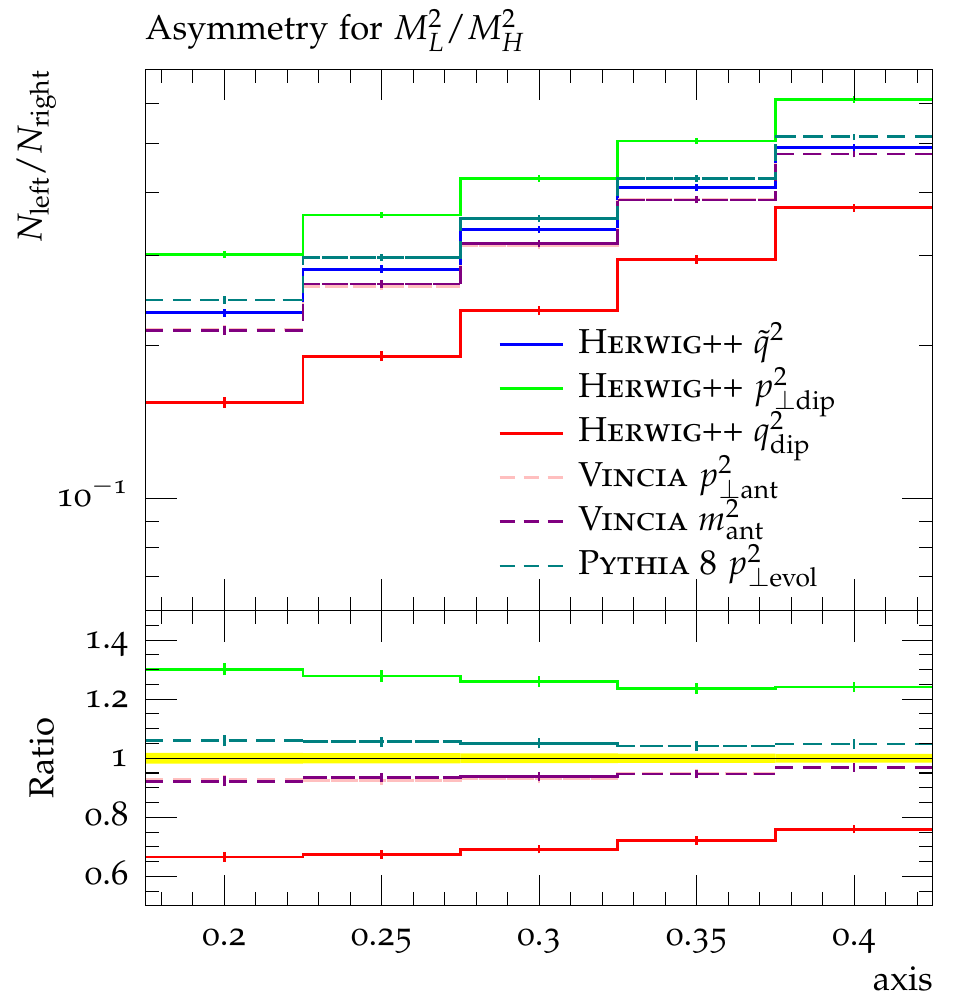}
\caption{The plots show the normalized distribution of the ratio of jet masses, 
$M_L^2/M_H^2$, on the left and its asymmetry with respect to the
  asymmetry axis, $x_0$, 
 on the right.}
\label{fig:ResObs3}
\end{figure}

To force a ``compressed'' scale hierarchy, we impose the cut $y_{34} >
0.5\, y_{23}$, and plot the ratio $M_L^2/M_H^2$ of the invariant
masses (squared) of the
jets at the end of the clustering, ordered so that $M_L^2 \le M^2_H$. 
With four partons at LO, the light
jet mass, $M_L$, is zero if both the $4\to3$ and $3\to2$ clusterings happen in
the same jet, while it is non-zero otherwise. Thus, the region close to zero
is sensitive to events with a $1\to 3$ splitting occurring in one
of the jets, while the region above $\sim 0.25$ is dominated by
opposite-side $1\to 2$ splittings.

The normalized distribution of the mass ratio is shown on the left side 
in \FigRef{fig:ResObs3}. The ratio plot shows that the difference between
the theory models mainly occurs in the region for values
$M_L^2/M_H^2\lesssim 0.3$, 
leaving smaller differences per bin for higher values due to the normalization.
To condense these difference we use the asymmetry, as defined in 
\EqRef{eq:asymmetry}, whose values are shown on the right in
\FigRef{fig:ResObs3} 
with respect to the asymmetry axis, $x_0$. The asymmetry roughly reflects the 
relative amount of events with a $1\to 3$ splitting occurring in one
of the jets, divided by events with opposite-side $1\to2$ splittings.

Compared to the angular-ordered shower of \tsc{Herwig++}, the 
$p^2_{\perp\mrm{dip}}$-ordered dipole shower and \tsc{Pythia~8} predict a higher
value for the asymmetry, whereas the the predictions of \tsc{Vincia}
with both ordering variables and the $q_\mrm{dip}^2$-ordered dipole 
shower of \tsc{Herwig++} are smaller. Both evolution variables of \tsc{Vincia}
result in the same value for the asymmetry, whereas a difference of $5\%$
to $12\%$ occurs between \tsc{Pythia~8} and \tsc{Vincia}. 
For \tsc{Pythia~8} and the $p^2_{\perp\mrm{ant}}$-ordered shower of \tsc{Vincia}
we also see differences up to nearly $20\%$ in the bins for small mass ratios.
Thus, this observable can be used to tell these theory models apart.

Note that we obtain a distribution similar to the mass ratio by using
the ECF, defined in \EqRef{eq:ECF}. The 1-point double ratio is defined as
\begin{align} \label{eq:C1beta}
C_1^{(\beta)}=\frac{\sum\limits_{i<j\in J}E_iE_j\theta_{ij}^\beta}{\Big(\sum\limits_{i\in J}E_i\Big)^2}~,
\end{align}
where the sum runs over all particles of a jet. By building a ratio
similar to the mass ratio we get
\begin{align} \label{eq:C1betaRatio}
\frac{C_{1,L}^{(\beta)}}{C_{1,H}^{(\beta)}} =
\frac{\sum\limits_{i<j\in J_L}E_iE_j\theta_{ij}^\beta}{\Big(\sum\limits_{i\in J_L}E_i\Big)^2}
\cdot 
\frac{\Big(\sum\limits_{i\in J_H}E_i\Big)^2}{\sum\limits_{i<j\in J_H}E_iE_j\theta_{ij}^\beta}~,
\end{align}
with $C_{1,L}^{(\beta)}\le C_{1,H}^{(\beta)}$.
With the expansion $\cos\theta\approx 1-\theta^2/2$, the invariant mass squared
of two particles is
\begin{align}
M_{ij}^2=E_iE_j(1-\cos\theta_{ij}) &\approx E_iE_j\theta_{ij}^2/2~.
\end{align}
By using a value of $\beta=2$, Eq.~\eqref{eq:C1betaRatio} is approximately equal
to the mass ratio and we thus obtain similar results for the two variables.

\section{Conclusions \label{sec:conclusions}}

We have studied four event-shape variables, designed to be 
sensitive to subleading aspects of the event structure in 
semi-inclusive $e^+e^- \to 4$-jet events, with a cut on $y_{34} >
0.0045$. 
Six different parton-shower
models were compared, available through the \tsc{Herwig++}, \tsc{Pythia~8},
and \tsc{Vincia} Monte Carlo codes. These models span a wide range of
theoretical ideas, 
from conventional parton showers to ones based on dipoles and
antennae, with different ordering criteria, different recoil
strategies, and different radiation functions. 

To make the comparison as fair and unbiased as possible, we first 
tuned all the theory models to the same set of existing LEP
measurements, using the \tsc{Professor} and \tsc{Rivet} tuning
tools. We find that the existing data already provides some
discriminating power, with the models using string hadronization
achieving somewhat lower $\chi^2$ values than those based on 
cluster hadronization\footnote{As emphasized in the main body of the
  paper, however, the unfolding of the data was based on the string
  model, with the cluster model only used to evaluate systematics, so
  there may be a slight bias towards favouring the string model
  inherent in the correction procedure, at a level at or below the
  hadronization component of the systematic uncertainty.}.
Therefore, it is important that we limit ourselves to draw
conclusions only from observables that are not very sensitive to
non--perturbative effects. 
\textsc{Vincia} with ordering in transverse
momentum provides the best overall description of the LEP data. Using just the
existing data, however, it is nearly impossible to tell, e.g.,
\tsc{Pythia} and \tsc{Vincia} apart, despite significant differences
between the shower models. Although the \tsc{Herwig++} models are easier to tell 
apart already using the existing data, we do also see larger differences 
between them in the variables proposed here, corroborating our conclusion 
that the new observables add significant discriminating power.

We have shown that the observables proposed here, which are sensitive
to coherence properties and to effective $1\to3$ splittings, allow for
additional significant discriminating power 
between the different models, given a sample size of order 500k events
or more.
The theory models for the shower implemented in \tsc{Herwig++} can clearly
be told apart by most of the observables we propose. Depending on the
tuning parameters, however, 
the cluster model may generate rather large non-perturbative
corrections to the 4-jet rate, especially at low $\theta_{14}$. For the
$\theta_{14}$ variable, we therefore also highlighted the integrated 
``Central/Away''
ratio as an observable that should be particularly robust against 
corrections at low $\theta_{14}$. 
As expected and measured with the 4-jet angular observable $\theta_{14}$, 
we see that the $m_{\mrm{ant}}^2$-ordered shower of \tsc{Vincia}
predicts a higher ratio of wide-angle to collinear emissions from a
three-jet system, compared to the $p^2_{\perp\mrm{ant}}$-ordered shower.
With the same observables as well as with the ratio of jet masses,
$M_L^2/M_H^2$, 
we can distinguish between \tsc{Vincia} and \tsc{Pythia~8}. The shower model of 
the latter produces more events where one hard jets lies back-to-back to the 
remaining jets of the event. 

We round off by emphasizing that a comparison against corrected LEP
data would be extremely interesting, and, we believe, 
of great importance to constraining the subleading properties of
modern-day QCD models. 

\subsection*{Acknowledgements}
We thank Stefan Kluth and Christoph Pahl for discussions on the 
experimental feasibility of the studies proposed here, and
Gavin Salam and Andrew Larkoski for pointing us to the ECF observables
and for useful discussions about the latter. NF would like to thank CERN
for hospitality during the course of this work. This work was supported by
the FP7 Marie Curie Initial Training Network MCnetITN under contract
PITN-GA-2012-315877.  SG and SP acknowledge support from
the Helmholtz Alliance ``Physics at the Terascale''. 

\appendix

\section{Tuning Observables, Weights and Parameters}

\begin{table}[tbh]
\begin{center}
\begin{tabular}{lr}
\toprule
Observable & Weight  \\ 
\midrule
$K^{*\pm}(892)$ spectrum & 1.0 \\
$\rho$ spectrum & 1.0 \\
$\omega(782)$ spectrum & 1.0 \\
$\Xi^-$ spectrum & 1.0 \\
$K^{*0}$ spectrum & 1.0 \\
$\phi$ spectrum & 1.0 \\
$\Sigma^\pm$ spectrum & 1.0 \\
$\gamma$ spectrum & 1.0 \\
$K^\pm$ spectrum & 1.0 \\
\bottomrule
\end{tabular} \hspace*{5mm}
\begin{tabular}{lr}
\toprule
Observable & Weight  \\ 
\midrule
$\Lambda^0$ spectrum & 1.0 \\
$\pi^0$ spectrum & 1.0 \\
$p$ spectrum & 1.0 \\
$\eta'$ spectrum & 1.0 \\
$\Xi^0(1530)$ spectrum & 1.0 \\
$\pi^\pm$ spectrum & 1.0 \\
$\eta$ spectrum & 1.0 \\
$K^0$ spectrum & 1.0 \\
 & \\
\bottomrule
\end{tabular}
\end{center}
\caption{Identified particle spectra and the associated weights, 
taken from Ref.~\cite{Abreu:1996na}.}
\label{tab:TuneObs_PS}
\end{table}

\begin{table}[tbh]
\begin{center}
\begin{tabular}{lr}
\toprule
Observable & Weight \\
\midrule
In-plane $p_\bot$ in GeV w.r.t. sphericity axes & 1.0 \\
In-plane $p_\bot$ in GeV w.r.t. thrust axes & 1.0 \\
Out-of-plane $p_\bot$ in GeV w.r.t. sphericity axes  & 1.0 \\
Out-of-plane $p_\bot$ in GeV w.r.t. thrust axes & 1.0 \\
Mean out-of-plane $p_\bot$ in GeV w.r.t. thrust axis vs. $x_p$ & 1.0 \\
Mean $p_\bot$ in GeV vs. $x_p$ & 1.0 \\
Scaled momentum $x_p=|p|/|p_\text{beam}|$ & 1.0 \\
Log of scaled momentum, $\log(1/x_p)$ & 1.0 \\
Energy-energy correlation, EEC & 1.0 \\
Sphericity, $S$ & 1.0 \\
Aplanarity, $A$ & 2.0 \\
Planarity, $P$ & 1.0 \\
$D$ parameter & 1.0 \\
$C$ parameter & 1.0 \\
1-Thrust & 1.0 \\
Thrust major, $M$ & 1.0 \\
Thrust minor, $m$ & 2.0\\
Oblatness, $O=M-m$ & 1.0 \\
Charged multiplicity distribution  & 2.0 \\
Mean charged multiplicity & 150.0 \\
\bottomrule
\end{tabular}
\end{center}
\caption{Event shapes and the associated weights, 
taken from Ref.~\cite{Abreu:1996na} and \cite{Barate:1996fi}.}
\label{tab:TuneObs_ES}
\end{table}

\begin{table}[tbh]
\begin{center}
\begin{tabular}{lr}
\toprule
Observable & Weight \\ 
\midrule
Differential 2-jet rate & 2.0 \\
Differential 3-jet rate & 2.0 \\
\bottomrule
\end{tabular} \hspace*{2mm}
\begin{tabular}{lr}
\toprule
Observable & Weight \\ 
\midrule
Differential 4-jet rate & 2.0 \\
Differential 5-jet rate & 2.0 \\
\bottomrule
\end{tabular}
\end{center}
\caption{Jet rates and the associated weights,
 taken from Ref.~\cite{Pfeifenschneider:1999rz}.}
\label{tab:TuneObs_JR}
\end{table}

\begin{table}[tbh]
\begin{center}
\begin{tabular}{lr}
\toprule
Observable & Weight \\ 
\midrule
$b$ quark fragmentation function $f(x_B^\text{weak})$ & 7.0 \\
Mean of $b$ quark fragmentation function $f(x_B^\text{weak})$ & 3.0 \\
\bottomrule
\end{tabular}
\end{center}
\caption{Observables for $b$ quarks and the associated weights, 
taken from Ref.~\cite{Heister:2001jg}.}
\label{tab:TuneObs_B}
\end{table}

\begin{table}[tbh]
\begin{center}
\begin{tabular}{lr}
\toprule
Observable & Weight  \\ 
\midrule
Mean $\rho^0(770)$ multiplicity & 10.0 \\
Mean $\Delta^{++}(1232)$ multiplicity & 10.0 \\
Mean $K^{*+}(892)$ multiplicity & 10.0 \\
Mean $\Sigma^0$ multiplicity & 10.0 \\
Mean $\Lambda_b^0$ multiplicity & 10.0 \\
Mean $K^+$ multiplicity & 10.0 \\
Mean $\Xi^0(1530)$ multiplicity & 10.0 \\
Mean $\Lambda(1520)$ multiplicity & 10.0 \\
Mean $D_s^{*+}(2112)$ multiplicity & 10.0 \\
Mean $\Sigma^-(1385)$ multiplicity & 10.0 \\
Mean $f_1(1420)$ multiplicity & 10.0 \\
Mean $\phi(1020)$ multiplicity & 10.0 \\
Mean $K_2^{*0}$ multiplicity & 10.0 \\
Mean $\Omega^-$ multiplicity & 10.0 \\
Mean $\Sigma^\pm(1385)$ multiplicity & 10.0 \\
Mean $\psi(2S)$ multiplicity & 10.0 \\
Mean $D^{*+}$ multiplicity & 10.0 \\
Mean $B^*$ multiplicity & 10.0 \\
Mean $\pi^0$ multiplicity & 10.0 \\
Mean $\eta$ multiplicity & 10.0 \\
Mean $a_0^+(980)$ multiplicity & 10.0 \\
Mean $D_{s1}^+$ multiplicity & 10.0 \\
Mean $\rho^+(770)$ multiplicity & 10.0 \\
Mean $\Xi^-$ multiplicity & 10.0 \\
Mean $\omega(782)$ multiplicity & 10.0 \\
Mean $\Upsilon(1S)$ multiplicity & 10.0 \\
\bottomrule
\end{tabular} \hspace*{5mm}
\begin{tabular}{lr}
\toprule
Observable & Weight  \\ 
\midrule
Mean $\chi_{c1}(3510)$ multiplicity & 10.0 \\
Mean $D^+$ multiplicity & 10.0 \\
Mean $\Sigma^+$ multiplicity & 10.0 \\
Mean $f_1(1285)$ multiplicity & 10.0 \\
Mean $f_2(1270)$ multiplicity & 10.0 \\
Mean $J/\psi(1S)$ multiplicity & 10.0 \\
Mean $B_u^+$ multiplicity & 10.0 \\
Mean $B^**$ multiplicity & 10.0 \\
Mean $\Lambda_c^+$ multiplicity & 10.0 \\
Mean $D^0$ multiplicity & 10.0 \\
Mean $f_2'(1525)$ multiplicity & 10.0 \\
Mean $\Sigma^\pm$ multiplicity & 10.0 \\
Mean $D_{s2}^+$ multiplicity & 10.0 \\
Mean $K^{*0}(892)$ multiplicity & 10.0 \\
Mean $\Sigma^-$ multiplicity & 10.0 \\
Mean $\pi^+$ multiplicity & 10.0 \\
Mean $f_0(980)$ multiplicity & 10.0 \\
Mean $\Sigma^+(1385)$ multiplicity & 10.0 \\
Mean $D_s^+$ multiplicity & 10.0 \\
Mean $p$ multiplicity & 10.0 \\
Mean $B_s^0$ multiplicity & 10.0 \\
Mean $K^0$ multiplicity & 10.0 \\
Mean $B^+,B_d^0$ multiplicity & 10.0 \\
Mean $\Lambda$ multiplicity & 10.0 \\
Mean $\eta'(958)$ multiplicity & 10.0 \\
 & \\
\bottomrule
\end{tabular}
\end{center}
\caption{Multiplicities and the associated weights, 
taken from Ref.~\cite{Amsler:2008zzb}.}
\label{tab:TuneObs_M}
\end{table}

\begin{table}[tp]
\begin{center}
\begin{tabular}{llll}
\toprule
 & \multicolumn{2}{l}{Parameter} & Description \\
\midrule
Default Shower & $\alpha_{M_Z}$ & \textsf{AlphaMZ} & Strong coupling at the $Z^0$ boson mass \\ 
 & $p_\text T^{\text{min}(f)}$ & \textsf{pTmin} & Shower cutoff \\
Dipole Shower & $\alpha_{M_Z}$ & \textsf{AlphaMZ} & Strong coupling at the $Z^0$ boson mass \\ 
 & $\mu_{\text{IR},FF}$ & \textsf{IRCutoff} & Infrared cutoff for final-final dipoles \\
 & $\mu^{(f)}_{\text{soft},FF}$ & \textsf{ScreeningScale} & Soft scale for final-final dipoles \\
Hadronization & $m_{g,c}$ & \textsf{ConsituentMass} & Gluon mass \\ 
 & $\text{Cl}^{(f)}_\text{max}$ & \textsf{ClMax} & Maximum cluster mass \\
 & $\text{Cl}^{(f)}_\text{pow}$ & \textsf{ClPow} & Cluster mass exponent \\
 & $\text{Cl}^{(f)}_\text{smr}$ & \textsf{ClSmr} & Smearing parameter \\
 & $P^{(f)}_\text{split}$ & \textsf{PSplit} & Mass exponent for daughter clusters \\
\bottomrule
\end{tabular}
\end{center}
\caption{The table lists the parameters for the \tsc{Herwig++} shower and hadronization model. 
The shower parameters indicated by the superscript $(f)$ exist in different copies for 
different splitting processes and the hadronization with superscript $(f)$ exist in
three copies for the different flavours: $(f)=(u,d,s),c,b$.}
\label{tab:TuneParamsH++} \vspace*{5mm}
\begin{center}
\begin{tabular}{llll}
\toprule
 & \multicolumn{2}{l}{Parameter} & Description \\
\midrule
Shower & $\alpha_{S}$ & \textsf{alphaS} & Strong coupling at the $Z^0$ boson mass \\ 
 & $p_\perp^{2\,\text{min}}$ & \textsf{cutoffScale} & Shower cutoff \\
Hadronization & $a_\text L$ & \textsf{aLund} & Parameter of the Lund symmetric fragmentation function \\
 & $b_\text L$ & \textsf{bLund} & Parameter of the Lund symmetric fragmentation function \\
 & $a_\text{ED}$ & \textsf{aExtraDiquark} & $a$ parameter for diquarks, with 
total $a=a_\text L+a_\text{ED}$ \\
 & $\sigma$ & \textsf{PTsigma} & Total width of the fragmentation $p_\bot$ \\
\bottomrule
\end{tabular}
\end{center}
\caption{The table lists the parameters for the shower model of \tsc{Vincia} and \tsc{Pythia~8} and
for the Lund hadronization model of \tsc{Pythia~8}.}
\label{tab:TuneParamsPV}
\end{table}

\begin{table}[tp]
\begin{center}
\begin{tabular}{lllllll}
\toprule
 & \multicolumn{2}{l}{Default Values} & & \multicolumn{3}{l}{Best Tune} \\
 &  &  &  & Default & Dipole & Dipole \\
Parameter & Default & Dipole & Range & $\tilde q^2$ & $p^2_{\perp\mrm{dip}}$ & $q_\mrm{dip}^2$ \\
\midrule
$\alpha_{M_Z}$ & $0.120$ & & $0.100-0.125$ & $0.123$ & & \\
$p_\text T^{\text{min}}$ & $1.00~\text{GeV}$ &  & $(0.50-1.50)~\text{GeV}$ & $1.39~\text{GeV}$ & & \\
$\alpha_{M_Z}$ & & $0.113$ & $0.100-0.138$ & & $0.128$ & $0.138$ \\
$\mu_{\text{IR},FF}$ &   & $1.41~\text{GeV}$ & $(0.50-2.00)~\text{GeV}$ & & $0.78~\text{GeV}$ & $0.72~\text{GeV}$ \\
$\mu_{\text{soft},FF}$ &  & $0.24~\text{GeV}$ & fixed & & $0.00~\text{GeV}$ & $0.00~\text{GeV}$ \\
$m_{g,c}$ & $0.95~\text{GeV}$ & $1.08~\text{GeV}$ & $(0.67-3.00)~\text{GeV}$
 & $0.70~\text{GeV}$ & $0.70~\text{GeV}$ & $0.96~\text{GeV}$\\
$\text{Cl}_\text{max}$ & $3.25~\text{GeV}$ & $4.17~\text{GeV}$ & $(2.00-4.50)~\text{GeV}$
 & $3.59~\text{GeV}$ & $3.12~\text{GeV}$ & $2.73~\text{GeV}$ \\
$\text{Cl}_\text{pow}$ & $1.28$ & $5.73$ & $2.00-10.00$ & $2.59$ & $5.72$ & $2.00$ \\
$\text{Cl}_\text{smr}$ & $0.78$ & $4.55$ & fixed & $0.78$ & $4.55$ & $4.55$ \\
$P_\text{split}$ & $1.14$ & $0.77$ & $0.00-1.40$ & $0.60$ & $0.74$ & $1.33$ \\
\bottomrule
\end{tabular}
\end{center} 
\caption{The table lists the parameters with their default value and the scanned range for 
the tuning of \tsc{Herwig++}. The last columns contain the values of the best tune.}
\label{tab:TuneResultsH++} \vspace*{5mm}
\begin{center}
\begin{tabular}{lllllll}
\toprule
 & \multicolumn{2}{l}{Default Values} & & \multicolumn{3}{l}{Best Tune} \\
 &  &  &  & \tsc{Pythia~8} & \tsc{Vincia} & \tsc{Vincia} \\
Parameter & \tsc{Pythia~8} & \tsc{Vincia} & Range & $p^2_{\perp\mrm{evol}}$ & $p_{\perp\mrm{ant}}^2$ & $m_{\mrm{ant}}^2$ \\
\midrule
$\alpha_{S}$ & $0.138$ & & $0.120-0.139$ & $0.139$ & & \\
$p_{\perp\mrm{evol}}^{2\,\text{min}}$ & $0.40$ & & $0.40-1.00$ & $0.41$ & & \\
$\alpha_{S}$ & & $0.129$ & $0.120-0.132$ & & $0.129$ & $0.132$ \\
$p_{\perp\mrm{ant}}^{2\,\text{min}}$ & & $0.60$ & $0.46-1.00$ & & $0.50$ & $0.76$ \\
$a_\text L$ & $0.30$ & $0.38$ & $0.20-0.70$ & $0.35$ & $0.38$ & $0.39$ \\
$b_\text L$ & $0.80$ & $0.90$ & $0.50-1.50$ & $0.94$ & $0.86$ & $0.71$ \\
$a_\text{ED}$ & $0.50$ & $1.00$ & $0.50-0.10$ & $0.95$ & $0.60$ & $0.55$ \\
$\sigma$ & $0.304$ & $0.275$ & $0.200-0.400$ & $0.284$ & $0.264$ & $0.291$ \\
\bottomrule
\end{tabular}
\end{center}
\caption{The table lists the parameters with their default value and the scanned range for 
the tuning of \tsc{Pythia~8} and \tsc{Vincia}. The last columns contain the values of the best tune.}
\label{tab:TuneResultsPV}
\end{table}


\begin{thebibliography}{10}

\bibitem{Buckley:2011ms}
{\bfseries MCnet} Collaboration, A.~Buckley {\em et~al.}, ``{General-purpose
  event generators for LHC physics},''
  \href{http://dx.doi.org/10.1016/j.physrep.2011.03.005}{{\em Phys.Rept.}
  {\bfseries 504} (2011) 145--233}.

\bibitem{Beringer:1900zz}
{\bfseries Particle Data Group} Collaboration, J.~Beringer {\em et~al.},
  ``{Review of Particle Physics (RPP)},''
\href{http://dx.doi.org/10.1103/PhysRevD.86.010001}{{\em Phys.Rev.} {\bfseries
  D86} (2012) 010001}.

\bibitem{Seymour:2013ega}
M.~H. Seymour and M.~Marx, ``{Monte Carlo Event Generators},''
\href{http://arxiv.org/abs/1304.6677}{{\ttfamily arXiv:1304.6677 [hep-ph]}}.

\bibitem{Gieseke:2013eva}
S.~Gieseke, ``{Simulation of jets at colliders},''
\href{http://dx.doi.org/10.1016/j.ppnp.2013.04.001}{{\em Prog.Part.Nucl.Phys.}
  {\bfseries 72} (2013) 155--205}.

\bibitem{Barate:1996fi}
{\bfseries ALEPH} Collaboration, R.~Barate {\em et~al.}, ``{Studies of quantum
  chromodynamics with the ALEPH detector},''
\href{http://dx.doi.org/10.1016/S0370-1573(97)00045-8}{{\em Phys.Rept.}
  {\bfseries 294} (1998) 1--165}.

\bibitem{Abreu:1996na}
{\bfseries DELPHI} Collaboration, P.~Abreu {\em et~al.}, ``{Tuning and test of
  fragmentation models based on identified particles and precision event shape
  data},''
\href{http://dx.doi.org/10.1007/s002880050295}{{\em Z.Phys.} {\bfseries C73}
  (1996) 11--60}.

\bibitem{Achard:2004sv}
{\bfseries L3} Collaboration, P.~Achard {\em et~al.}, ``{Studies of hadronic
  event structure in $e^{+} e^{-}$ annihilation from 30-GeV to 209-GeV with the
  L3 detector},'' \href{http://dx.doi.org/10.1016/j.physrep.2004.07.002}{{\em
  Phys.Rept.} {\bfseries 399} (2004) 71--174},
\href{http://arxiv.org/abs/hep-ex/0406049}{{\ttfamily arXiv:hep-ex/0406049
  [hep-ex]}}.

\bibitem{Akrawy:1990yx}
{\bfseries OPAL} Collaboration, M.~Akrawy {\em et~al.}, ``{A Measurement of
  Global Event Shape Distributions in the Hadronic Decays of the {$Z^0$}},''
\href{http://dx.doi.org/10.1007/BF01552315}{{\em Z.Phys.} {\bfseries C47}
  (1990) 505--522}.

\bibitem{Corcella:2000bw}
G.~Corcella, I.~Knowles, G.~Marchesini, S.~Moretti, K.~Odagiri, {\em et~al.},
  ``{HERWIG 6: An Event generator for hadron emission reactions with
  interfering gluons (including supersymmetric processes)},''
  \href{http://dx.doi.org/10.1088/1126-6708/2001/01/010}{{\em JHEP} {\bfseries
  0101} (2001) 010},
\href{http://arxiv.org/abs/hep-ph/0011363}{{\ttfamily arXiv:hep-ph/0011363
  [hep-ph]}}.

\bibitem{Sjostrand:2006za}
T.~Sj{\"o}strand, S.~Mrenna, and P.~Z. Skands, ``{PYTHIA 6.4 Physics and
  Manual},'' \href{http://dx.doi.org/10.1088/1126-6708/2006/05/026}{{\em JHEP}
  {\bfseries 0605} (2006) 026},
\href{http://arxiv.org/abs/hep-ph/0603175}{{\ttfamily arXiv:hep-ph/0603175
  [hep-ph]}}.

\bibitem{Lonnblad:1992tz}
L.~L{\"o}nnblad, ``{ARIADNE version 4: A Program for simulation of QCD cascades
  implementing the color dipole model},''
\href{http://dx.doi.org/10.1016/0010-4655(92)90068-A}{{\em Comput.Phys.Commun.}
  {\bfseries 71} (1992) 15--31}.

\bibitem{Buckley:2010ar}
{\bfseries RIVET} Collaboration, A.~Buckley, J.~Butterworth, L.~L{\"o}nnblad,
  H.~Hoeth, J.~Monk, {\em et~al.}, ``{Rivet user manual},''
  \href{http://arxiv.org/abs/1003.0694}{{\ttfamily arXiv:1003.0694 [hep-ph]}}.

\bibitem{Abreu:1990ce}
{\bfseries DELPHI} Collaboration, P.~Abreu {\em et~al.}, ``{Experimental study
  of the triple gluon vertex},''
\href{http://dx.doi.org/10.1016/0370-2693(91)90796-S}{{\em Phys.Lett.}
  {\bfseries B255} (1991) 466--476}.

\bibitem{Decamp:1992ip}
{\bfseries ALEPH} Collaboration, D.~Decamp {\em et~al.}, ``{Evidence for the
  triple gluon vertex from measurements of the QCD color factors in Z decay
  into four jets},''
\href{http://dx.doi.org/10.1016/0370-2693(92)91941-2}{{\em Phys.Lett.}
  {\bfseries B284} (1992) 151--162}.

\bibitem{Abbiendi:2001qn}
{\bfseries OPAL} Collaboration, G.~Abbiendi {\em et~al.}, ``{A Simultaneous
  measurement of the QCD color factors and the strong coupling},''
  \href{http://dx.doi.org/10.1007/s100520100699}{{\em Eur.Phys.J.} {\bfseries
  C20} (2001) 601--615},
\href{http://arxiv.org/abs/hep-ex/0101044}{{\ttfamily arXiv:hep-ex/0101044
  [hep-ex]}}.

\bibitem{Abdallah:2004uu}
{\bfseries DELPHI} Collaboration, J.~Abdallah {\em et~al.}, ``{Coherent soft
  particle production in Z decays into three jets},''
  \href{http://dx.doi.org/10.1016/j.physletb.2004.10.059}{{\em Phys.Lett.}
  {\bfseries B605} (2005) 37--48},
\href{http://arxiv.org/abs/hep-ex/0410075}{{\ttfamily arXiv:hep-ex/0410075
  [hep-ex]}}.

\bibitem{Abbiendi:2003ri}
{\bfseries OPAL} Collaboration, G.~Abbiendi {\em et~al.}, ``{Tests of models of
  color reconnection and a search for glueballs using gluon jets with a
  rapidity gap},'' \href{http://dx.doi.org/10.1140/epjc/s2004-01809-2}{{\em
  Eur.Phys.J.} {\bfseries C35} (2004) 293--312},
\href{http://arxiv.org/abs/hep-ex/0306021}{{\ttfamily arXiv:hep-ex/0306021
  [hep-ex]}}.

\bibitem{Achard:2003pe}
{\bfseries L3} Collaboration, P.~Achard {\em et~al.}, ``{Search for color
  reconnection effects in $e^{+} e^{-} \to W^{+} W^{-} \to$ hadrons through
  particle flow studies at LEP},''
  \href{http://dx.doi.org/10.1016/S0370-2693(03)00490-8}{{\em Phys.Lett.}
  {\bfseries B561} (2003) 202--212},
\href{http://arxiv.org/abs/hep-ex/0303042}{{\ttfamily arXiv:hep-ex/0303042
  [hep-ex]}}.

\bibitem{Achard:2003ik}
{\bfseries L3} Collaboration, P.~Achard {\em et~al.}, ``{Search for color
  singlet and color reconnection effects in hadronic $Z$ decays at LEP},''
  \href{http://dx.doi.org/10.1016/j.physletb.2003.12.003}{{\em Phys.Lett.}
  {\bfseries B581} (2004) 19--30},
\href{http://arxiv.org/abs/hep-ex/0312026}{{\ttfamily arXiv:hep-ex/0312026
  [hep-ex]}}.

\bibitem{Siebel:2005uw}
{\bfseries DELPHI} Collaboration, M.~Siebel, ``{A Study of the charge of
  leading hadrons in gluon and quark fragmentation},''
\href{http://arxiv.org/abs/hep-ex/0505080}{{\ttfamily arXiv:hep-ex/0505080
  [hep-ex]}}.

\bibitem{Abbiendi:2005es}
{\bfseries OPAL} Collaboration, G.~Abbiendi {\em et~al.}, ``{Colour
  reconnection in $e^+e^- \to W^+ W^-$ at s**(1/2) = 189-GeV - 209-GeV},''
  \href{http://dx.doi.org/10.1140/epjc/s2005-02439-x}{{\em Eur.Phys.J.}
  {\bfseries C45} (2006) 291--305},
\href{http://arxiv.org/abs/hep-ex/0508062}{{\ttfamily arXiv:hep-ex/0508062
  [hep-ex]}}.

\bibitem{Schael:2006ns}
{\bfseries ALEPH} Collaboration, S.~Schael {\em et~al.}, ``{Test of Colour
  Reconnection Models using Three-Jet Events in Hadronic Z Decays},''
  \href{http://dx.doi.org/10.1140/epjc/s10052-006-0017-5}{{\em Eur.Phys.J.}
  {\bfseries C48} (2006) 685--698},
\href{http://arxiv.org/abs/hep-ex/0604042}{{\ttfamily arXiv:hep-ex/0604042
  [hep-ex]}}.

\bibitem{Abdallah:2006uq}
{\bfseries DELPHI} Collaboration, J.~Abdallah {\em et~al.}, ``{Investigation of
  colour reconnection in WW events with the DELPHI detector at LEP-2},''
  \href{http://dx.doi.org/10.1140/epjc/s10052-007-0304-9}{{\em Eur.Phys.J.}
  {\bfseries C51} (2007) 249--269},
\href{http://arxiv.org/abs/0704.0597}{{\ttfamily arXiv:0704.0597 [hep-ex]}}.

\bibitem{Karneyeu:2013aha}
A.~Karneyeu, L.~Mijovic, S.~Prestel, and P.~Skands, ``{MCPLOTS: a particle
  physics resource based on volunteer computing},''
  \href{http://arxiv.org/abs/1306.3436}{{\ttfamily arXiv:1306.3436 [hep-ph]}}.
\url{http://mcplots.cern.ch}.

\bibitem{Gieseke:2003rz}
S.~Gieseke, P.~Stephens, and B.~Webber, ``{New formalism for QCD parton
  showers},'' {\em JHEP} {\bfseries 0312} (2003) 045,
\href{http://arxiv.org/abs/hep-ph/0310083}{{\ttfamily arXiv:hep-ph/0310083
  [hep-ph]}}.

\bibitem{Sjostrand:2004ef}
T.~Sj{\"o}strand and P.~Z. Skands, ``{Transverse-momentum-ordered showers and
  interleaved multiple interactions},''
  \href{http://dx.doi.org/10.1140/epjc/s2004-02084-y}{{\em Eur.Phys.J.}
  {\bfseries C39} (2005) 129--154},
\href{http://arxiv.org/abs/hep-ph/0408302}{{\ttfamily arXiv:hep-ph/0408302
  [hep-ph]}}.

\bibitem{Nagy:2005aa}
Z.~Nagy and D.~E. Soper, ``{Matching parton showers to NLO computations},''
  \href{http://dx.doi.org/10.1088/1126-6708/2005/10/024}{{\em JHEP} {\bfseries
  0510} (2005) 024},
\href{http://arxiv.org/abs/hep-ph/0503053}{{\ttfamily arXiv:hep-ph/0503053
  [hep-ph]}}.

\bibitem{Krauss:2005re}
F.~Krauss, A.~Schalicke, and G.~Soff, ``{APACIC++ 2.0: A Parton cascade in
  C++},'' \href{http://dx.doi.org/10.1016/j.cpc.2005.11.009}{{\em
  Comput.Phys.Commun.} {\bfseries 174} (2006) 876--902},
\href{http://arxiv.org/abs/hep-ph/0503087}{{\ttfamily arXiv:hep-ph/0503087
  [hep-ph]}}.

\bibitem{Giele:2007di}
W.~T. Giele, D.~A. Kosower, and P.~Z. Skands, ``{A simple shower and matching
  algorithm},'' \href{http://dx.doi.org/10.1103/PhysRevD.78.014026}{{\em
  Phys.Rev.} {\bfseries D78} (2008) 014026},
\href{http://arxiv.org/abs/0707.3652}{{\ttfamily arXiv:0707.3652 [hep-ph]}}.

\bibitem{Dinsdale:2007mf}
M.~Dinsdale, M.~Ternick, and S.~Weinzierl, ``{Parton showers from the dipole
  formalism},'' \href{http://dx.doi.org/10.1103/PhysRevD.76.094003}{{\em
  Phys.Rev.} {\bfseries D76} (2007) 094003},
\href{http://arxiv.org/abs/0709.1026}{{\ttfamily arXiv:0709.1026 [hep-ph]}}.

\bibitem{Platzer:2009jq}
S.~Pl{\"a}tzer and S.~Gieseke, ``{Coherent Parton Showers with Local
  Recoils},'' \href{http://dx.doi.org/10.1007/JHEP01(2011)024}{{\em JHEP}
  {\bfseries 1101} (2011) 024},
\href{http://arxiv.org/abs/0909.5593}{{\ttfamily arXiv:0909.5593 [hep-ph]}}.

\bibitem{Nagy:2012bt}
Z.~Nagy and D.~E. Soper, ``{Parton shower evolution with subleading color},''
  \href{http://dx.doi.org/10.1007/JHEP06(2012)044}{{\em JHEP} {\bfseries 1206}
  (2012) 044},
\href{http://arxiv.org/abs/1202.4496}{{\ttfamily arXiv:1202.4496 [hep-ph]}}.

\bibitem{Schumann:2007mg}
S.~Schumann and F.~Krauss, ``{A Parton shower algorithm based on Catani-Seymour
  dipole factorisation},''
  \href{http://dx.doi.org/10.1088/1126-6708/2008/03/038}{{\em JHEP} {\bfseries
  0803} (2008) 038},
\href{http://arxiv.org/abs/0709.1027}{{\ttfamily arXiv:0709.1027 [hep-ph]}}.

\bibitem{Winter:2007ye}
J.-C. Winter and F.~Krauss, ``{Initial-state showering based on colour dipoles
  connected to incoming parton lines},''
  \href{http://dx.doi.org/10.1088/1126-6708/2008/07/040}{{\em JHEP} {\bfseries
  0807} (2008) 040},
\href{http://arxiv.org/abs/0712.3913}{{\ttfamily arXiv:0712.3913 [hep-ph]}}.

\bibitem{Bahr:2008pv}
M.~B{\"a}hr, S.~Gieseke, M.~Gigg, D.~Grellscheid, K.~Hamilton, {\em et~al.},
  ``{Herwig++ Physics and Manual},''
  \href{http://dx.doi.org/10.1140/epjc/s10052-008-0798-9}{{\em Eur.Phys.J.}
  {\bfseries C58} (2008) 639--707},
\href{http://arxiv.org/abs/0803.0883}{{\ttfamily arXiv:0803.0883 [hep-ph]}}.

\bibitem{Sjostrand:2007gs}
T.~Sj{\"o}strand, S.~Mrenna, and P.~Z. Skands, ``{A Brief Introduction to
  PYTHIA 8.1},'' \href{http://dx.doi.org/10.1016/j.cpc.2008.01.036}{{\em
  Comput.Phys.Commun.} {\bfseries 178} (2008) 852--867},
\href{http://arxiv.org/abs/0710.3820}{{\ttfamily arXiv:0710.3820 [hep-ph]}}.

\bibitem{Gleisberg:2008ta}
T.~Gleisberg, S.~H{\"o}che, F.~Krauss, M.~Sch{\"o}nherr, S.~Schumann, {\em
  et~al.}, ``{Event generation with SHERPA 1.1},''
  \href{http://dx.doi.org/10.1088/1126-6708/2009/02/007}{{\em JHEP} {\bfseries
  0902} (2009) 007},
\href{http://arxiv.org/abs/0811.4622}{{\ttfamily arXiv:0811.4622 [hep-ph]}}.

\bibitem{Gustafson:1987rq}
G.~Gustafson and U.~Pettersson, ``{Dipole Formulation of QCD Cascades},''
\href{http://dx.doi.org/10.1016/0550-3213(88)90441-5}{{\em Nucl.Phys.}
  {\bfseries B306} (1988) 746}.

\bibitem{Catani:1996vz}
S.~Catani and M.~Seymour, ``{A General algorithm for calculating jet
  cross-sections in NLO QCD},''
  \href{http://dx.doi.org/10.1016/S0550-3213(96)00589-5}{{\em Nucl.Phys.}
  {\bfseries B485} (1997) 291--419},
\href{http://arxiv.org/abs/hep-ph/9605323}{{\ttfamily arXiv:hep-ph/9605323
  [hep-ph]}}.

\bibitem{Kosower:2003bh}
D.~A. Kosower, ``{Antenna factorization in strongly ordered limits},''
  \href{http://dx.doi.org/10.1103/PhysRevD.71.045016}{{\em Phys.Rev.}
  {\bfseries D71} (2005) 045016},
\href{http://arxiv.org/abs/hep-ph/0311272}{{\ttfamily arXiv:hep-ph/0311272
  [hep-ph]}}.

\bibitem{GehrmannDeRidder:2005cm}
A.~Gehrmann-De~Ridder, T.~Gehrmann, and E.~N. Glover, ``{Antenna subtraction at
  NNLO},'' \href{http://dx.doi.org/10.1088/1126-6708/2005/09/056}{{\em JHEP}
  {\bfseries 0509} (2005) 056},
\href{http://arxiv.org/abs/hep-ph/0505111}{{\ttfamily arXiv:hep-ph/0505111
  [hep-ph]}}.

\bibitem{Platzer:2011bc}
S.~Pl{\"a}tzer and S.~Gieseke, ``{Dipole Showers and Automated NLO Matching in
  Herwig++},'' \href{http://dx.doi.org/10.1140/epjc/s10052-012-2187-7}{{\em
  Eur.Phys.J.} {\bfseries C72} (2012) 2187},
\href{http://arxiv.org/abs/1109.6256}{{\ttfamily arXiv:1109.6256 [hep-ph]}}.

\bibitem{Giele:2011cb}
W.~Giele, D.~Kosower, and P.~Skands, ``{Higher-Order Corrections to Timelike
  Jets},'' \href{http://dx.doi.org/10.1103/PhysRevD.84.054003}{{\em Phys.Rev.}
  {\bfseries D84} (2011) 054003},
\href{http://arxiv.org/abs/1102.2126}{{\ttfamily arXiv:1102.2126 [hep-ph]}}.

\bibitem{Hartgring:2013jma}
L.~Hartgring, E.~Laenen, and P.~Skands, ``{Antenna Showers with One-Loop Matrix
  Elements},''
\href{http://arxiv.org/abs/1303.4974}{{\ttfamily arXiv:1303.4974 [hep-ph]}}.

\bibitem{AlcarazMaestre:2012vp}
J.~Alcaraz~Maestre {\em et~al.}, ``{The SM and NLO Multileg and SM MC Working
  Groups: Summary Report},''
\href{http://arxiv.org/abs/1203.6803}{{\ttfamily arXiv:1203.6803 [hep-ph]}}.

\bibitem{Larkoski:2013eya}
A.~J. Larkoski, G.~P. Salam, and J.~Thaler, ``{Energy Correlation Functions for
  Jet Substructure},'' \href{http://dx.doi.org/10.1007/JHEP06(2013)108}{{\em
  JHEP} {\bfseries 1306} (2013) 108},
\href{http://arxiv.org/abs/1305.0007}{{\ttfamily arXiv:1305.0007 [hep-ph]}}.

\bibitem{Gribov:1972ri}
V.~N. Gribov and L.~N. Lipatov, ``{Deep inelastic $e$-$p$ scattering in
  perturbation theory},'' {\em Sov. J. Nucl. Phys.} {\bfseries 15} (1972) 438.

\bibitem{Altarelli:1977zs}
G.~Altarelli and G.~Parisi, ``{Asymptotic freedom in parton language},'' {\em
  Nucl. Phys.} {\bfseries B126} (1977) 298.

\bibitem{Dokshitzer:1977sg}
Y.~L. Dokshitzer, ``{Calculation of the structure functions for deep inelastic
  scattering and $e^+e^-$ annihilation by perturbation theory in quantum
  chromodynamics},'' {\em Sov. Phys. JETP} {\bfseries 46} (1977) 641.

\bibitem{Bengtsson:1986et}
M.~Bengtsson and T.~Sj{\"o}strand, ``{A Comparative Study of Coherent and
  Noncoherent Parton Shower Evolution},''
\href{http://dx.doi.org/10.1016/0550-3213(87)90407-X}{{\em Nucl.Phys.}
  {\bfseries B289} (1987) 810}.

\bibitem{Marchesini:1983bm}
G.~Marchesini and B.~Webber, ``{Simulation of QCD Jets Including Soft Gluon
  Interference},''
\href{http://dx.doi.org/10.1016/0550-3213(84)90463-2}{{\em Nucl.Phys.}
  {\bfseries B238} (1984) 1}.

\bibitem{nadine}
N.~Fischer, ``{Angular Correlations and Soft Jets as Probes of Parton
  Showers},'' Master's thesis, KIT, 2013.

\bibitem{Buckley:2009bj}
{\bfseries Professor} Collaboration, A.~Buckley, H.~Hoeth, H.~Lacker,
  H.~Schulz, and J.~E. von Seggern, ``{Systematic event generator tuning for
  the LHC},'' \href{http://dx.doi.org/10.1140/epjc/s10052-009-1196-7}{{\em
  Eur.Phys.J.} {\bfseries C65} (2010) 331},
\href{http://arxiv.org/abs/0907.2973}{{\ttfamily arXiv:0907.2973 [hep-ph]}}.

\bibitem{Heister:2001jg}
{\bfseries ALEPH} Collaboration, A.~Heister {\em et~al.}, ``{Study of the
  fragmentation of b quarks into B mesons at the Z peak},''
  \href{http://dx.doi.org/10.1016/S0370-2693(01)00690-6}{{\em Phys.Lett.}
  {\bfseries B512} (2001) 30},
\href{http://arxiv.org/abs/hep-ex/0106051}{{\ttfamily arXiv:hep-ex/0106051
  [hep-ex]}}.

\bibitem{Pfeifenschneider:1999rz}
{\bfseries JADE, OPAL} Collaboration, P.~Pfeifenschneider {\em et~al.}, ``{QCD
  analyses and determinations of alpha(s) in e+ e- annihilation at energies
  between 35-GeV and 189-GeV},''
  \href{http://dx.doi.org/10.1007/s100520000432}{{\em Eur.Phys.J.} {\bfseries
  C17} (2000) 19},
\href{http://arxiv.org/abs/hep-ex/0001055}{{\ttfamily arXiv:hep-ex/0001055
  [hep-ex]}}.

\bibitem{Amsler:2008zzb}
{\bfseries Particle Data Group} Collaboration, C.~Amsler {\em et~al.},
  ``{Review of Particle Physics},''
\href{http://dx.doi.org/10.1016/j.physletb.2008.07.018}{{\em Phys.Lett.}
  {\bfseries B667} (2008) 1--1340}.

\bibitem{Sjostrand:1993hi}
T.~Sj{\"o}strand and V.~A. Khoze, ``{On Color rearrangement in hadronic W+ W-
  events},'' \href{http://dx.doi.org/10.1007/BF01560244}{{\em Z.Phys.}
  {\bfseries C62} (1994) 281--310},
\href{http://arxiv.org/abs/hep-ph/9310242}{{\ttfamily arXiv:hep-ph/9310242
  [hep-ph]}}.

\bibitem{Rathsman:1998tp}
J.~Rathsman, ``{A Generalized area law for hadronic string re-interactions},''
  \href{http://dx.doi.org/10.1016/S0370-2693(99)00291-9}{{\em Phys.Lett.}
  {\bfseries B452} (1999) 364--371},
\href{http://arxiv.org/abs/hep-ph/9812423}{{\ttfamily arXiv:hep-ph/9812423
  [hep-ph]}}.

\bibitem{Skands:2007zg}
P.~Z. Skands and D.~Wicke, ``{Non-perturbative QCD effects and the top mass at
  the Tevatron},'' \href{http://dx.doi.org/10.1140/epjc/s10052-007-0352-1}{{\em
  Eur.Phys.J.} {\bfseries C52} (2007) 133--140},
\href{http://arxiv.org/abs/hep-ph/0703081}{{\ttfamily arXiv:hep-ph/0703081
  [HEP-PH]}}.

\bibitem{Platzer:2012np}
S.~Pl{\"a}tzer and M.~Sj{\"o}dahl, ``{Subleading $N_c$ improved Parton
  Showers},'' \href{http://dx.doi.org/10.1007/JHEP07(2012)042}{{\em JHEP}
  {\bfseries 1207} (2012) 042},
\href{http://arxiv.org/abs/1201.0260}{{\ttfamily arXiv:1201.0260 [hep-ph]}}.

\bibitem{Gieseke:2012ft}
S.~Gieseke, C.~R{\"o}hr, and A.~Siodmok, ``{Colour reconnections in
  Herwig++},'' \href{http://dx.doi.org/10.1140/epjc/s10052-012-2225-5}{{\em
  Eur.Phys.J.} {\bfseries C72} (2012) 2225},
\href{http://arxiv.org/abs/1206.0041}{{\ttfamily arXiv:1206.0041 [hep-ph]}}.

\bibitem{Sjostrand:1993yb}
T.~Sj{\"o}strand, ``{High-energy physics event generation with PYTHIA 5.7 and
  JETSET 7.4},''
\href{http://dx.doi.org/10.1016/0010-4655(94)90132-5}{{\em Comput.Phys.Commun.}
  {\bfseries 82} (1994) 74--90}.

\bibitem{Catani:1991hj}
S.~Catani, Y.~L. Dokshitzer, M.~Olsson, G.~Turnock, and B.~Webber, ``{New
  clustering algorithm for multi - jet cross-sections in e+ e- annihilation},''
\href{http://dx.doi.org/10.1016/0370-2693(91)90196-W}{{\em Phys.Lett.}
  {\bfseries B269} (1991) 432--438}.

\end{thebibliography}
\providecommand{\href}[2]{#2}\begingroup\raggedright\endgroup

\end{document}